\documentstyle[12pt,epsf]{article}

\catcode`@=11
\def\chkspace{%
  \relax   
  \begingroup\ifhmode\aftergroup\dochksp@ce\fi\endgroup}
\def\dochksp@ce{%
  \unskip              
  \futurelet\chkspct@k\d@chkspc  
}
\def\d@chkspc{%
  \let\nxtsp@ce=\relax
  \ifx\chkspct@k.\else     
    \ifx\chkspct@k,\else
      \ifx\chkspct@k;\else
        \ifx\chkspct@k!\else
          \ifx\chkspct@k?\else
            \ifx\chkspct@k:\else
              \ifx\chkspct@k)\else
              \ifx\chkspct@k(\else
                \ifx\chkspct@k]\else
                  \ifx\chkspct@k-\else
                    \ifx\chkspct@k\egroup\else  
                      \let\nxtsp@ce=\put@space  
                    \fi
                  \fi
                \fi
              \fi
              \fi
            \fi
          \fi
        \fi
      \fi
    \fi
  \fi
  \nxtsp@ce
}
\def\put@space{$\;$}
\catcode`@=12

\def\ra{{$\rightarrow$}\chkspace}

\def\etal{{\it et al.}\chkspace}

\def\adhoc{{\it ad hoc}\chkspace}

\def\eg{{\it eg.}\chkspace}

\def\ep{{e$^+$e$^-$}\chkspace}

\def\gluino{\relax\ifmmode \tilde{g} \else $\tilde{g}$ \fi\chkspace}

\chkspace
\chkspace
\def\m0{$M_{0}$}\chkspace
\def\m0m{$M_{0}MAX$}\chkspace
\chkspace
\chkspace

\def\bbrm{\relax\ifmmode {\rm b}\bar{\rm b}
       \else ${\rm b}\bar{\rm b}$ \fi\chkspace}
\def\bb{$b\bar{b}$ \chkspace}
\def\ccrm{\relax\ifmmode {\rm c}\bar{\rm c}
       \else ${\rm c}\bar{\rm c}$ \fi\chkspace}
\def\cc{$c\bar{c}$ \chkspace}
\def\tt{\relax\ifmmode {\rm t}\bar{\rm t}
       \else ${\rm t}\bar{\rm t}$ \fi\chkspace}
\def\ss{\relax\ifmmode {\rm s}\bar{\rm s}
       \else ${\rm s}\bar{\rm s}$ \fi\chkspace}
\def\uu{\relax\ifmmode {\rm u}\bar{\rm u}
       \else ${\rm u}\bar{\rm u}$ \fi\chkspace}
\def\dd{\relax\ifmmode {\rm d}\bar{\rm d}
       \else ${\rm d}\bar{\rm d}$ \fi\chkspace}

\def\qqg{\relax\ifmmode {\rm q}\overline{\rm q}{\rm g}
\else q$\overline{\rm q}$g \fi\chkspace}

\def\afb{\relax\ifmmode A_{FB} \else
{{$A_{FB}$}}\fi\chkspace}
\def\afbb{\relax\ifmmode A_{FB}^b \else
{{$A_{FB}^b$}}\fi\chkspace}
\def\pafb{\relax\ifmmode \tilde{A}_{FB} \else
{{$\tilde{A}_{FB}$}}\fi\chkspace}
\def\pafbb{\relax\ifmmode \tilde{A}_{FB}^b \else
{{$\tilde{A}_{FB}^b$}}\fi\chkspace}

\def\pafbzo{\relax\ifmmode \tilde{A}_{FB}|_{O(0)} \else
{{$\tilde{A}_{FB}|_{O(0)}$}}\fi\chkspace}
\def\pafbfo{\relax\ifmmode \tilde{A}_{FB}|_{\oalp} \else
{{$\tilde{A}_{FB}|_{\oalp}$}}\fi\chkspace}
\def\pafbso{\relax\ifmmode \tilde{A}_{FB}|_{\oalpsq} \else
{{$\tilde{A}_{FB}|_{\oalpsq}$}}\fi\chkspace}
\def\pafbto{\relax\ifmmode \tilde{A}_{FB}|_{\oalpc} \else
{{$\tilde{A}_{FB}|_{\oalpc}$}}\fi\chkspace}

\def\pafbbzo{\relax\ifmmode \tilde{A}_{FB}^b|_{O(0)} \else
{{$\tilde{A}_{FB}^b|_{O(0)}$}}\fi\chkspace}
\def\pafbbfo{\relax\ifmmode \tilde{A}_{FB}^b|_{\oalp} \else
{{$\tilde{A}_{FB}^b|_{\oalp}$}}\fi\chkspace}
\def\pafbbso{\relax\ifmmode \tilde{A}_{FB}^b|_{\oalpsq} \else
{{$\tilde{A}_{FB}^b|_{\oalpsq}$}}\fi\chkspace}
\def\pafbbto{\relax\ifmmode \tilde{A}_{FB}^b|_{\oalpc} \else
{{$\tilde{A}_{FB}^b|_{\oalpc}$}}\fi\chkspace}

\def\afbo0{\tilde{A}_{FB}|_{O(0)}}
\def\afbo1{\tilde{A}_{FB}|_{\oalp}}
\def\afbo2{\tilde{A}_{FB}|_{\oalpsq}}
\def\afbo3{\tilde{A}_{FB}|_{\oalpc}}

\def\lam{\relax\ifmmode \Lambda_{\overline{MS}}
       \else {{$\Lambda_{\overline{MS}}$}}\fi\chkspace}
\def\lamuds{\relax\ifmmode \Lambda^{(3)}_{\overline{MS}}
       \else {{$\Lambda^{(3)}_{\overline{MS}}$}}\fi\chkspace}
\def\lamudsc{\relax\ifmmode \Lambda^{(4)}_{\overline{MS}}
       \else $\Lambda^{(4)}_{\overline{MS}}$\fi\chkspace}
\def\lamudscb{\relax\ifmmode \Lambda^{(5)}_{\overline{MS}}
       \else $\Lambda^{(5)}_{\overline{MS}}$\fi\chkspace}

\def\alp{\relax\ifmmode \alpha_s\else $\alpha_s$\fi\chkspace}
\def\alpbar{\relax\ifmmode \bar{\alpha_s}
       \else $\bar{\alpha_s}$\fi\chkspace}
\def\alpmz{\relax\ifmmode \alpha_s(M_Z)\else $\alpha_s(M_Z)$\fi\chkspace}
\def\alpmzsq{\relax\ifmmode \alpha_s(M_Z^2)
       \else $\alpha_s(M_Z^2)$\fi\chkspace}

\def\oalp{\relax\ifmmode O(\alpha_s)\else{{O($\alpha_s$)}}\fi\chkspace}
\def\oalpsq{\relax\ifmmode O(\alpha_s^2)
           \else{{O($\alpha_s^2$)}}\fi\chkspace}
\def\oalpc{\relax\ifmmode O(\alpha_s^3)
           \else{{O($\alpha_s^3$)}}\fi\chkspace}
\def\oalpf{\relax\ifmmode O(\alpha_s^4)
           \else{{O($\alpha_s^4$)}}\fi\chkspace}

\def\rb{\relax\ifmmode R_3^b/R_3^{all}
           \else{{$R_3^b/R_3^{all}$}}\fi\chkspace}
\def\rc{\relax\ifmmode R_3^c/R_3^{all}
           \else{{$R_3^c/R_3^{all}$}}\fi\chkspace}
\def\ruds{\relax\ifmmode R_3^{uds}/R_3^{all}
           \else{{$R_3^{uds}/R_3^{all}$}}\fi\chkspace}
\def\ri{\relax\ifmmode R_3^i/R_3^{all}
           \else{{$R_3^i/R_3^{all}$}}\fi\chkspace}
\def\rj{\relax\ifmmode R_3^j/R_3^{all}
           \else{{$R_3^j/R_3^{all}$}}\fi\chkspace}
\def\alpi{\relax\ifmmode \alpha^i_s/\alpha^{all}_s
           \else{{$\alpha^i_s/\alpha^{all}_s$}}\fi\chkspace}

\def\plb{Phys. Lett.\chkspace}
\def\npb{Nucl. Phys.\chkspace}

\def\prl{Phys. Rev. Lett.\chkspace}
\def\prd{Phys. Rev.\chkspace}
\def\zpc{Z. Phys.\chkspace}

\def\z0{{$Z^0$}\chkspace}
\def\Dst{\relax\ifmmode {\rm D}^* \else {D$^*$}\fi\chkspace}
\def\Dpl{\relax\ifmmode {\rm D}^+ \else {D$^+$}\fi\chkspace}
\def\D0{\relax\ifmmode {\rm D}^0 \else {D$^0$}\fi\chkspace}
\def\Kst{\relax\ifmmode {\rm K}^* \else {K$^*$}\fi\chkspace}
\def\K0{\relax\ifmmode {\rm K}^0_s \else {K$^0_s$}\fi\chkspace}
\def\Kpl{\relax\ifmmode {\rm K}^+ \else {K$^+$}\fi\chkspace}
\def\Kstz{\relax\ifmmode {\rm K}^{*0} \else {K$^{*0}$}\fi\chkspace}

\def\ep{{$e^+e^-$}\chkspace}

\def\z0{$Z^0$}
\def\xb{{$x_B$}\chkspace}
\def\bb{{$b\bar{b}$}\chkspace}
\def\cc{{$c\bar{c}$}\chkspace}

\def\etal{{\it et al.}\chkspace}
\def\adhoc{{\it ad hoc}\chkspace}

\def\qqg{{$q\bar{q}g$}\chkspace}

\renewcommand{\baselinestretch}{1.0}

 1

\catcode`\@=11 
\makeatletter
\def\@seccntformat#1{\csname the#1\endcsname.\hskip 1em}
\makeatother

\begin{document}

\thispagestyle{empty}
\begin{flushright}
{\renewcommand{\baselinestretch}{.75}
  SLAC--PUB--8153\\
June 1999\\
}
\end{flushright}

\vskip 1truecm
 
\begin{center}
{\large\bf
 Measurement of the $b$ Quark Fragmentation \\ Function in $Z^0$ decays$^*$
}
\end{center}

\begin{center}
 {\bf The SLD Collaboration$^{**}$}\\
Stanford Linear Accelerator Center \\
Stanford University, Stanford, CA~94309
\end{center}
 
\vspace{1cm}
 
\begin{center}
{\bf ABSTRACT }
\end{center}

\noindent
We present preliminary results of a new measurement 
of the inclusive $b$ quark fragmentation function in $Z^{0}$ 
decays using a novel kinematic $B$ hadron energy 
reconstruction technique.  The measurement is performed
using 150,000 hadronic \z0 events recorded in the SLD experiment 
at SLAC between 1996 and 1997.  The small and stable SLC beam spot 
and the CCD-based vertex detector are used to reconstruct 
topological $B$-decay vertices with high efficiency and purity, 
and to provide precise measurements of the kinematic
quantities used in this technique.  We measure the $B$ energy 
with good efficiency and resolution over the full kinematic range.  
We compare the measured scaled $B$ hadron energy distribution with several 
functional forms of the $B$ hadron energy distribution and predictions 
of several models of $b$ quark fragmentation.  Several functions 
are excluded by the data.  The average scaled energy of the weakly 
decaying $B$ hadron is measured to be 
\xb $=$ 0.714 $\pm$ 0.005 (stat) $\pm$ 0.007 (syst) 
$\pm$ 0.002 (model) (preliminary).

\vspace {1.4cm}
 
\vfill
\noindent
Contributed to: the International Europhysics Conference on High Energy Physics,
15-21 July 1999, Tampere, Finland, Ref. 1\_185, and to the XIXth International 
Symposium on Lepton and Photon Interactions, August 9-14 1999, Stanford, USA.

{\footnotesize
$^*$ Work supported by Department of Energy contract DE-AC03-76SF00515 (SLAC).}

\eject

%

%

\eject
\rm  
\section{Introduction}
\noindent 
The production of heavy hadrons (H) in \ep annihilation provides a
laboratory for the study of heavy-quark (Q) jet fragmentation. This is 
commonly characterised in terms of the observable 
$x_{H}$ $\equiv$ $2E_H/\sqrt{s}$, where
$E_H$ is the energy of a $B$ or $D$ hadron containing a $b$ or $c$ quark,
respectively, and $\sqrt{s}$ is the c.m. energy. In contrast to light-quark
jet fragmentation one expects~\cite{Bj} the distribution of $x_{H}$, 
$D(x_{H})$, to peak at an $x_{H}$-value significantly above 0. 
Since the hadronisation process is intrinsically non-perturbative $D(x_{H})$ 
cannot be calculated directly using perturbative Quantum Chromodynamics
(QCD). However, the distribution of the closely-related variable
$x_{Q}$ $\equiv$ 2$E_Q/\sqrt{s}$ can be calculated
perturbatively \cite{mn,dkt,bcfy} and related, via model-dependent
assumptions, to the observable quantity $D(x_{H})$; a number of such
models of heavy quark fragmentation have been proposed
\cite{lund,bowler,pete}. Measurements of $D(x_{H})$ thus serve to
constrain both perturbative QCD and the model predictions. 
Furthermore, the measurement of $D(x_{H})$ at different c.m. energies
can be used to test QCD evolution, and comparison of $D(x_{B})$
with $D(x_{D})$ can be used to test heavy quark symmetry~\cite{jaffe,Lisa}. 
Finally, the uncertainty on the forms of $D(x_{D})$ and $D(x_{B})$
must be taken into account in studies of the production and decay of heavy
quarks, see \eg~\cite{heavy}; more accurate measurements of these forms 
will allow increased precision in tests of the electroweak heavy-quark sector.

We consider the measurement of the $B$ hadron scaled energy distribution
$D(x_{B})$ in $Z^0$ decays. Earlier studies \cite{early} 
used the momentum spectrum of the lepton from semi-leptonic $B$ decays to 
constrain the mean value $<x_{B}>$ and found it to be approximately
$0.70$; this is in agreement with the results of similar studies at $\sqrt{s}$
= 29 and 35 GeV~\cite{petra}. In more recent
analyses~\cite{aleph95,shape,sldbfrag} 
the scaled energy distribution 
$D(x_{B})$ has been measured by reconstructing $B$ hadrons via their
$B$ \ra D$l$X decay mode. In this case the reconstruction efficiency is
intrinsically low due to the small branching ratio for $B$ hadrons to decay into
the high-momentum leptons used in the tag.  Also,
the reconstruction of the $B$ hadron energy using calorimeter information 
usually has poor resolution for low $B$ energy, resulting
in poor sensitivity to the shape of the distribution at low energy.

Here we describe the preliminary results of a new method for reconstructing
$B$ hadron decays, and the $B$ energy, inclusively, using only charged tracks,
in the SLD experiment at SLAC.
We use the upgraded CCD vertex detector, 
installed in 1996, to reconstruct $B$-decay vertices with high 
efficiency and purity.  Combined with the micron-size SLC interaction point
(IP), precise vertexing allows us to reconstruct accurately
the $B$ flight direction and hence 
the transverse momentum of tracks associated with 
the vertex with respect to this direction.  
Using the transverse momentum and 
the total invariant mass of the associated tracks, an upper limit
on the mass of the missing particles is found for each 
reconstructed $B$-decay vertex, and is used to solve for the longitudinal 
momentum of the missing particles, and hence for the energy 
of the $B$ hadron. In order 
to improve the $B$ sample purity and the reconstructed $B$ hadron energy 
resolution, $B$ vertices with low missing mass are selected.
The method is described in Section 3. In Section 4
we compare the $B$ energy distribution with predictions 
of heavy quark fragmentation models.  We also test several functional 
forms of $B$ hadron energy distributions.  In Section 5, we unfolded
the $B$ hadron energy distribution.  In Section 6, we discuss the 
systematic errors.  In Section 7 we summarize the results.
 
\section{Apparatus and Hadronic Event Selection}
 
\noindent
This analysis is based on roughly 150,000 hadronic events produced in 
\ep annihilations at a mean center-of-mass energy of $\sqrt{s}=91.28$ GeV
at the SLAC Linear Collider (SLC), and recorded in the SLC Large Detector
(SLD) in 1996 and 1997. 
A general description of the SLD can be found elsewhere~\cite{sld}.
The trigger and initial selection criteria for hadronic $Z^0$ decays are 
described in Ref.~\cite{sldalphas}.
This analysis used charged tracks measured in the Central Drift
Chamber (CDC)~\cite{cdc} and in the upgraded Vertex Detector (VXD3)~\cite{vxd}.
Momentum measurement is provided by a uniform axial magnetic field of 0.6T.
The CDC and VXD3  give a momentum resolution of
$\sigma_{p_{\perp}}/p_{\perp}$ = $0.01 \oplus 0.0026p_{\perp}$,
where $p_{\perp}$ is the track momentum transverse to the beam axis in
GeV/$c$. In the plane normal to the beamline 
the centroid of the micron-sized SLC IP is reconstructed from tracks
in sets of approximately thirty sequential hadronic \z0 decays to a precision 
of $\sigma^{r\phi}\simeq7\pm2$ $\mu$m (1996)
and $\sigma^{r\phi}\simeq4\pm2$ $\mu$m (1997).  The IP position along the 
beam axis is determined event by event using charged tracks with 
a resolution of $\sigma^z$ $\simeq$ 35 $\mu$m (1996) and 
 $\sigma^z$ $\simeq$ 30 $\mu$m (1997).  
Including the uncertainty on the IP position, the resolution on the 
charged-track impact parameter ($d$) projected in the plane perpendicular
to the beamline is 
$\sigma_{d}^{r\phi}$ = 14$\oplus$33/$(p\sin^{3/2}\theta)$ $\mu$m 
(1996) and
$\sigma_{d}^{r\phi}$ = 11$\oplus$33/$(p\sin^{3/2}\theta)$ $\mu$m 
(1997),
and the resolution in the plane containing the beam axis is  
$\sigma_{d}^{z}$ = 27$\oplus$33/$(p\sin^{3/2}\theta)$ $\mu$m
(1996) and
$\sigma_{d}^{z}$ = 24$\oplus$33/$(p\sin^{3/2}\theta)$ $\mu$m 
(1997),
where
$\theta$ is the track polar angle with respect to the beamline.
The event thrust axis~\cite{thrust} is calculated using energy clusters
measured in the Liquid Argon Calorimeter~\cite{lac}. 

A set of cuts is applied to the data to select well-measured tracks
and events well contained within the detector acceptance.
Charged tracks are required to have a distance of
closest approach transverse to the beam axis within 5 cm,
and within 10 cm along the axis from the measured IP,
as well as $|\cos \theta |< 0.80$, and $p_\perp > 0.15$ GeV/c.
Events are required to have a minimum of seven such tracks,
a thrust axis  polar angle w.r.t. the beamline, $\theta_T$,
within $|\cos\theta_T|<0.71$, and
a charged visible energy $E_{vis}$ of at least 20~GeV,
which is calculated from the selected tracks assigned the charged pion mass. 
The efficiency for selecting a well-contained $Z^0 \rightarrow q{\bar q}(g)$
event is estimated to be above 96\% independent of quark flavor. The
selected sample comprised 111,569 events, with an estimated
$0.10 \pm 0.05\%$ background contribution dominated
by $Z^0 \rightarrow \tau^+\tau^-$ events.

For the purpose of estimating the efficiency and purity of the $B$ hadron
selection procedure we made use of a detailed Monte Carlo (MC) simulation 
of the detector.
The JETSET 7.4~\cite{jetset} event generator is used, with parameter
values tuned to hadronic \ep annihilation data~\cite{tune},
combined with a simulation of $B$ hadron decays
tuned~\cite{sldsim} to $\Upsilon(4S)$ data and a simulation of the SLD
based on GEANT 3.21~\cite{geant}.
Inclusive distributions of single-particle and event-topology observables
in hadronic events are found to be well described by the
simulation~\cite{sldalphas}. Uncertainties in the simulation 
are taken into account in the systematic errors (Section~\ref{sec:sys}). 

\noindent
\section{$B$ Hadron Selection and Energy Measurement}

\subsection{$B$ Hadron Selection}

The $B$ sample for this analysis is selected using a topological vertexing
technique based on the detection and measurement of charged tracks,
which is described in detail in Ref.~\cite{zvnim}. 
Each hadronic event is divided into two hemispheres by a plane perpendicular
to the thrust axis.
In each hemisphere the topological vertexing algorithm is applied to 
the set of ` quality' tracks having
(i) at least 23 hits in the CDC and 2 hits in VXD3; 
(ii) a combined CDC and VXD3 track fit quality of $\chi^{2}/N_{dof}< $8;
(iii) a momentum in the range 0.25$<p<$55 GeV/$c$,
(iv) an impact parameter of less than 0.3~cm 
in the $r\phi$ plane, and less than 1.5~cm along the $z$ axis;
(v) a transverse impact parameter error no larger than 250 $\mu$m. 

Vertices consistent with photon conversions or $K^{0}$ and $\Lambda^0$ decays 
are discarded.
In hemispheres containing at least one found vertex the
vertex furthest from the IP is retained 
as the `seed' vertex.  
Those events are retained which contain a seed vertex separated from the IP 
by between
0.1~cm and 2.3~cm. The lower bound reduces contamination from non-$B$-decay
tracks and backgrounds from light-flavor events, and the upper bound
reduces the background from particle interactions with the beam 
pipe.

For each hemisphere containing an accepted seed vertex, a 
vertex axis is formed by the straight line joining the IP to
the seed vertex, which is located at a distance D from the IP.  
For each quality track not directly associated with the vertex,
the distance of closest approach to the vertex axis, T,
and the distance from the IP along the vertex
axis to the point of closest approach, L, are calculated. 
Tracks satisfying T$<1$~mm and L$/$D$>0.3$ are added to the vertex.
These T and L cuts are chosen to minimize false track associations 
to the seed vertex, since typically the addition of
a false track has a much greater
kinematic effect than the omission of a genuine $B$-decay track, and hence 
has more effect on the reconstructed $B$ hadron energy resolution.  
Our Monte Carlo studies show that, on average, this procedure 
attaches 0.85 tracks to each seed vertex, 91.9\% of the tracks 
from tagged true $B$ decays are associated
with the resulting vertices, and 98.0\% of the vertex tracks are from true
$B$ decays.  

The large masses of the $B$ hadrons relative to light-flavor hadrons 
make it possible to distinguish $B$ hadron decay vertices from those 
vertices found in events of light flavors using the vertex invariant 
mass, $M$. However, due to the missing particles, which are mainly neutrals,
$M$ cannot be fully determined.  
In the {\em rest} frame of the decaying hadron, $M$ can be written as 
\begin{equation}
M=\sqrt{M_{ch}^{2}+P_{t}^{2}+P_{chl}^{2}}+\sqrt{M_{0}^{2}+P_{t}^{2}
\label{eqn:vertexmass}
+P_{0l}^{2}}
\end{equation}
where $M_{ch}$ and $M_{0}$ are the total invariant masses of the set of 
vertex-associated tracks and the set of missing particles, respectively.
$P_{t}$ is the total charged track momentum transverse to the $B$ flight 
direction, which is identical to the transverse momentum of the set of 
missing particles by momentum conservation.  $P_{chl}$ and $P_{0l}$ are 
the respective momenta along the $B$ flight direction.  
In the $B$ {\em rest} frame, $P_{chl} = P_{0l}$.  
Using the set of vertex-associated charged tracks, we calculate
the total momentum vector ${\vec{P}}_{ch}$, the total energy $E_{ch}$ 
and the invariant mass $M_{ch}$, assuming the charged pion mass for each
track.
%
%
The $B$ hadron flight direction (the line joining the IP and the $B$ vertex.  
The lower bound for the mass of the decaying hadron, 
the `$P_{t}$-corrected vertex mass', 
\vspace{-0.2cm}
\begin{equation}
M_{Pt} = \sqrt{M_{ch}^{2}+P_{t}^{2}} + |P_{t}|
\label{eqn:masspt}
\end{equation}
is used as the variable for selecting $B$ hadrons.
The majority of non-$B$ vertices have $M_{Pt}$ less than 2.0 GeV/$c^{2}$.  
However, occasionally the measured $P_t$ may fluctuate to a 
much larger 
value than the true $P_t$, causing some charm vertices to have a $M_{Pt}$ 
larger than 2.0 GeV/$c^{2}$.  
To reduce this contamination, we calculate the `minimum $P_t$' by 
allowing the locations of the IP and the vertex to float to any pair 
of locations within the respective one sigma error-ellipsoids, 
We substitute the minimum $P_t$ in Equation~(\ref{eqn:masspt}) and 
use the modified $M_{Pt}$ as our 
variable for selecting $B$ hadrons~\cite{sldrb98}.  

\begin{figure}[ht]	
\centerline{\epsfxsize 1.1 truein 
\epsfysize6.5 in
\epsfxsize6.5 in
\leavevmode
\epsfbox{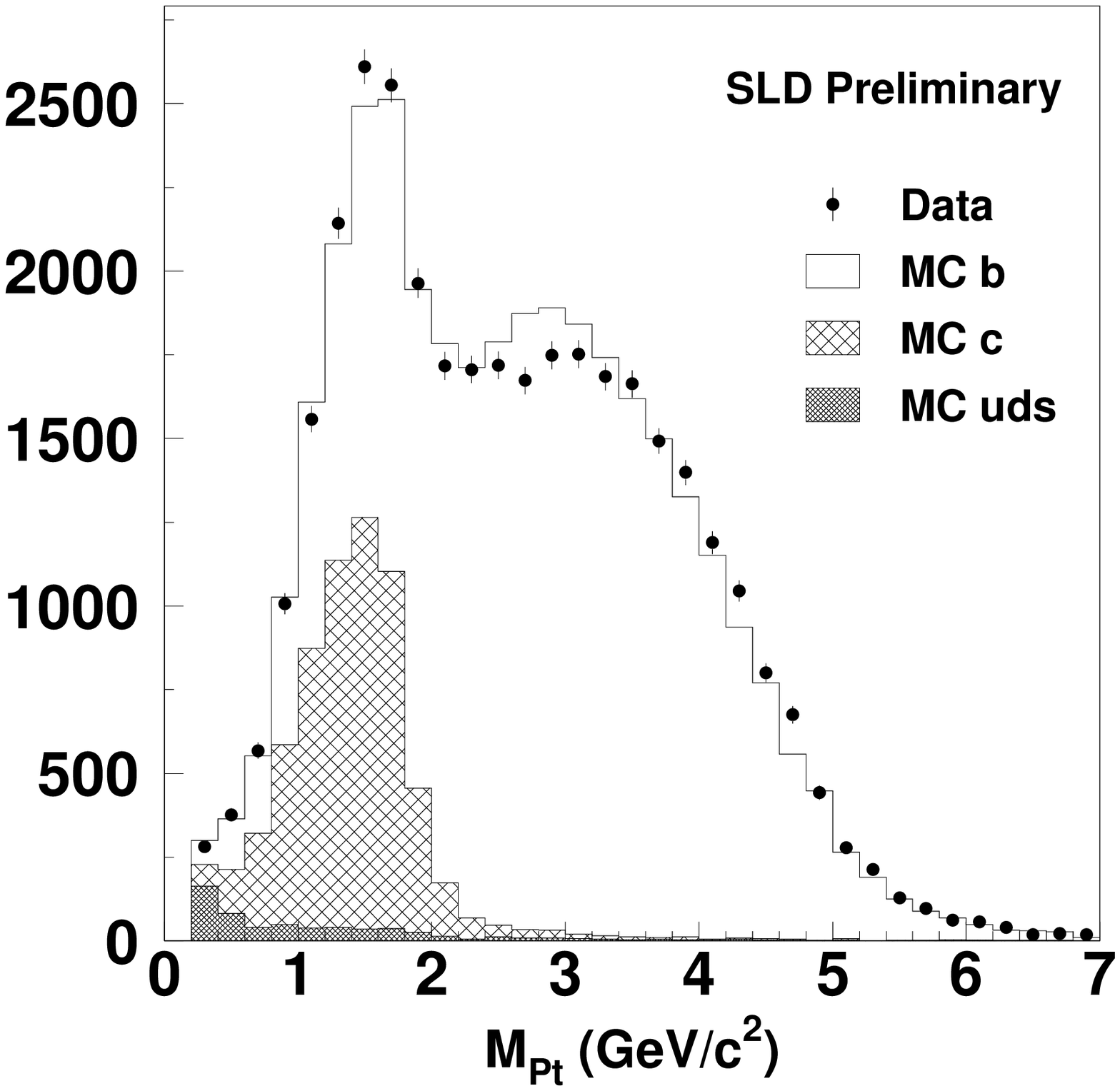}
}
\vskip -.2 cm
\caption[]{
\label{mptm}
Distribution of the reconstructed $P_{t}$-corrected vertex mass in 
the 1996-97 data (points).  Also shown is the prediction of the Monte 
Carlo simulation, for which the flavor composition is indicated.
}
\end{figure}

Figure~\ref{mptm} shows the distribution of the $M_{Pt}$ 
%
%
for the 32,492 hemispheres in the data sample with a found secondary 
vertex, and the corresponding simulated distribution (histogram).  
$B$ hadron candidates are selected by requiring 
$M_{Pt}$ $>$ 2.0 GeV/$c^{2}$.  We further required 
$M_{Pt} \leq 2 \times M_{ch}$ to reduce the contamination from fake
vertices in light quark events~\cite{sldrb98}.
A total of 19,604 hemispheres are selected, 
with an estimated efficiency for selecting a true $B$-hemisphere 
of 40.1\%, and a sample purity of 98.2\%.  The contributions from 
light-flavor events in the sample are 0.15\% for primary u,d and s events
and 1.6\% for c events.

\subsection{$B$ Hadron Energy Measurement}

The energy of each $B$ hadron, $E_{B}$, can be expressed as
the sum of the reconstructed-vertex energy, $E_{ch}$, 
and the energy of those particles not associated with the vertex, $E_{0}$.  

We can write $E_{0}$ as 
\begin{equation}
  E_{0}^{2} =  M_{0}^{2} + P_{t}^{2} + P_{0l}^{2} 
\label{eqn:e0}
\end{equation}
The two unknowns, $M_{0}$ and $P_{0l}$, must be found in order 
to obtain $E_{0}$.
One kinematic constraint can be obtained by imposing the $B$ hadron mass 
on the vertex, $M_{B}^{2}=E_{B}^{2}-P_{B}^{2}$, where 
$P_{B}=P_{chl}+P_{0l}$ is the total momentum of the $B$ hadron, 
and $P_{chl}$ is the momentum component of the vertex-associated
tracks along the vertex axis.  From Equation~(\ref{eqn:vertexmass}) we 
derive the following inequality,
\begin{equation}
  \sqrt{M_{ch}^2 + P_{t}^2} + \sqrt{M_{0}^2 + P_{t}^2} \leq M_{B}, 
\label{massineq}
\end{equation}
where equality holds in the limit where
both $P_{0l}$ and $P_{chl}$ vanish in the $B$ hadron {\em rest} frame.
Equation~(\ref{massineq}) effectively sets an upper bound on 
$M_{0}$, and a lower bound is given by zero:
\begin{equation}
   0\leq M_{0}^{2}\leq M_{0max}^{2},
\end{equation}
where 
\begin{equation}
M_{0max}^{2}=M_{B}^2 - 2M_{B}\sqrt{M_{ch}^2+P_{t}^2} + M_{ch}^2. 
\label{m0maxeqn}
\end{equation}
Since $M_{0}$ is bounded from both above and below, we expect to obtain
a good estimate of $M_{0}$, and therefore of the $B$ hadron energy,  
when $M_{0max}^{2}$ is small. 

\begin{figure}[ht]	
\centerline{\epsfxsize 1.1 truein 
\epsfysize6.5 in
\epsfxsize6.5 in
\leavevmode
\epsfbox{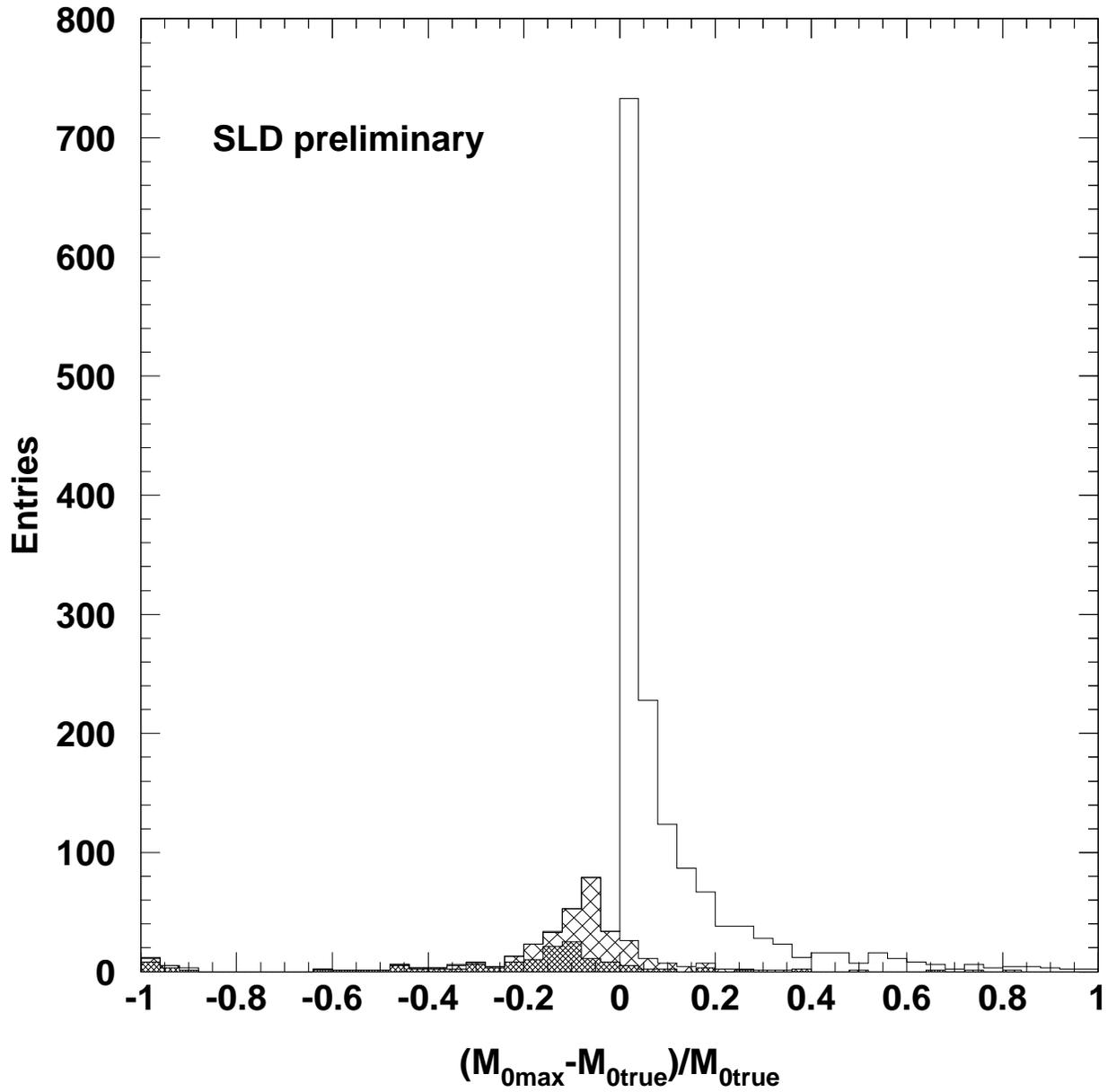}
}
\vskip -.2 cm
\caption[]{
\label{m0max_m0}
The relative deviation of the maximum missing mass from the true 
missing mass for Monte Carlo simulated $B$ hadron decays, which is divided
into three categories: $B^{0}$ and $B^{\pm}$ (open), 
$B_{s}^{0}$ (cross-hatched), and $\Lambda_{b}$ (dark hatched).
}
\end{figure}

\begin{figure}[ht]	
\centerline{\epsfxsize 1.1 truein 
\epsfysize6.5 in
\epsfxsize6.5 in
\leavevmode
\epsfbox{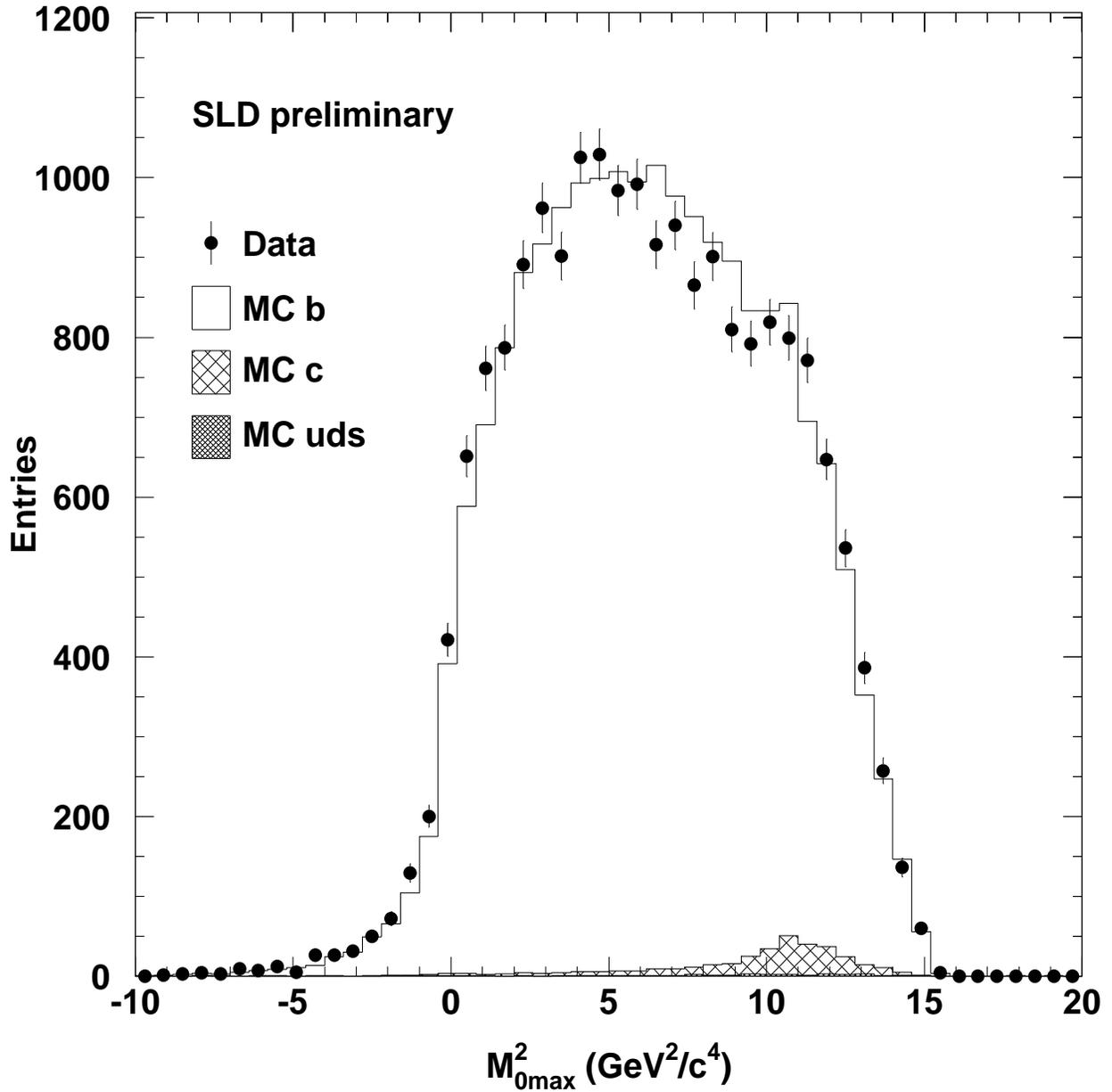}
}
\vskip -.2 cm
\caption[]{
\label{m0max_after}
Distribution of the reconstructed $M_{0max}^{2}$ for the selected 
vertices in the 1996-97 data (points).  
Also shown is the prediction of the Monte Carlo simulation.
}
\end{figure}

We have used our simulation to study this issue.
Assuming $M_{B}=$ 5.28 GeV/$c^{2}$, the true value of 
$M_{0}$ tends to cluster near its maximum value $M_{0max}$.
Figure~\ref{m0max_m0} shows 
the relative deviation of $M_{0max}$ from $M_{0true}$ for all $B$ hadrons.
Although approximately 20\% of the $B$ hadrons are $B^{0}_{s}$ and 
$\Lambda_{b}$ which have larger 
masses, the values of $M_{0max}$ obtained using $M_{B}$=5.28 GeV/$c^{2}$ 
in Equation~(\ref{m0maxeqn}) 
are typically within about 10\% of $M_0$.
The distribution of the reconstructed $M_{0max}^{2}$ for vertices in
the selected $B$ hadron sample is shown in Figure~\ref{m0max_after}.
The simulation indicates that the
non-$b\bar{b}$ background is concentrated at high $M_{0max}^{2}$; this because
most of the light flavor vertices have small $M_{Pt}$
and therefore, due to the strong negative correlation between 
$M_{Pt}$ and $M_{0max}$, 
large $M_{0max}$. The negative tail in Figure~\ref{m0max_after}
is an effect of detector resolution, and  
the Monte Carlo simulation shows good agreement with the data.

Because $M_{0}$ peaks near $M_{0max}$, 
we set $M_{0}^{2}$ = $M_{0max}^{2}$ if $M_{0max}^{2}$ $\geq$0, and
$M_{0}^{2}$ = 0 if $M_{0max}^{2}$ $<$0.
We then calculate $P_{0l}$:
\begin{equation}
   P_{0l} = \frac {\textstyle M_{B}^{2}-(M_{ch}^{2}+P_{t}^{2})-(M_{0}^{2}+P_{t}^{2})}{\textstyle 2 (M_{ch}^{2}+P_{t}^{2})} P_{chl},
\label{eqn:p0l} 
\end{equation}
and hence $E_{0}$ (Equation~(\ref{eqn:e0})).  
We then divide the reconstructed $B$ hadron energy, 
$E_{B}^{rec}=E_{0}+E_{ch}$, by the beam energy, $E_{beam}=\sqrt{s}/2$, 
to obtain the reconstructed scaled $B$ hadron energy, 
$x_{B}^{rec}=E_{B}^{rec}/E_{beam}$.

The resolution of $x_{B}^{rec}$ depends on both $M_{0max}^{2}$ 
and the true $x_{B}$, $x_{B}^{true}$.  Vertices in the negative tail of
the $M_{0max}^2$ distribution that have $M_{0max}^{2}<-1.0 (GeV/c^{2})^{2}$ 
are often poorly reconstructed and are not used in further analysis.
Vertices with small values of $|M_{0max}^{2}|$ are typically reconstructed
with better resolution and
an upper cut on $M_{0max}^{2}$ is hence applied.  
For an $x_B$-independent cut, 
the efficiency for selecting $B$ hadrons is roughly linear in $x_{B}^{true}$.
In order to obtain an approximately $x_B$-independent selection efficiency
we choose the following upper cut:
\begin{equation}
        M_{0max}^{2} < \left\{ 1.1+0.006 (E_{beam}-E_{B}^{rec})+
       3.5 exp[-(E_{B}^{rec}-5.5)/3.5] \right\}^2,
\label{eqn:m0maxcut} 
\end{equation}
where the two terms that depend on the reconstructed energy $E_{B}^{rec}$ 
increase the efficiency at lower $B$ hadron energy.

\begin{figure}[ht]	
\centerline{\epsfxsize 1.1 truein 
\epsfysize6.5 in
\epsfxsize6.5 in
\leavevmode
\epsfbox{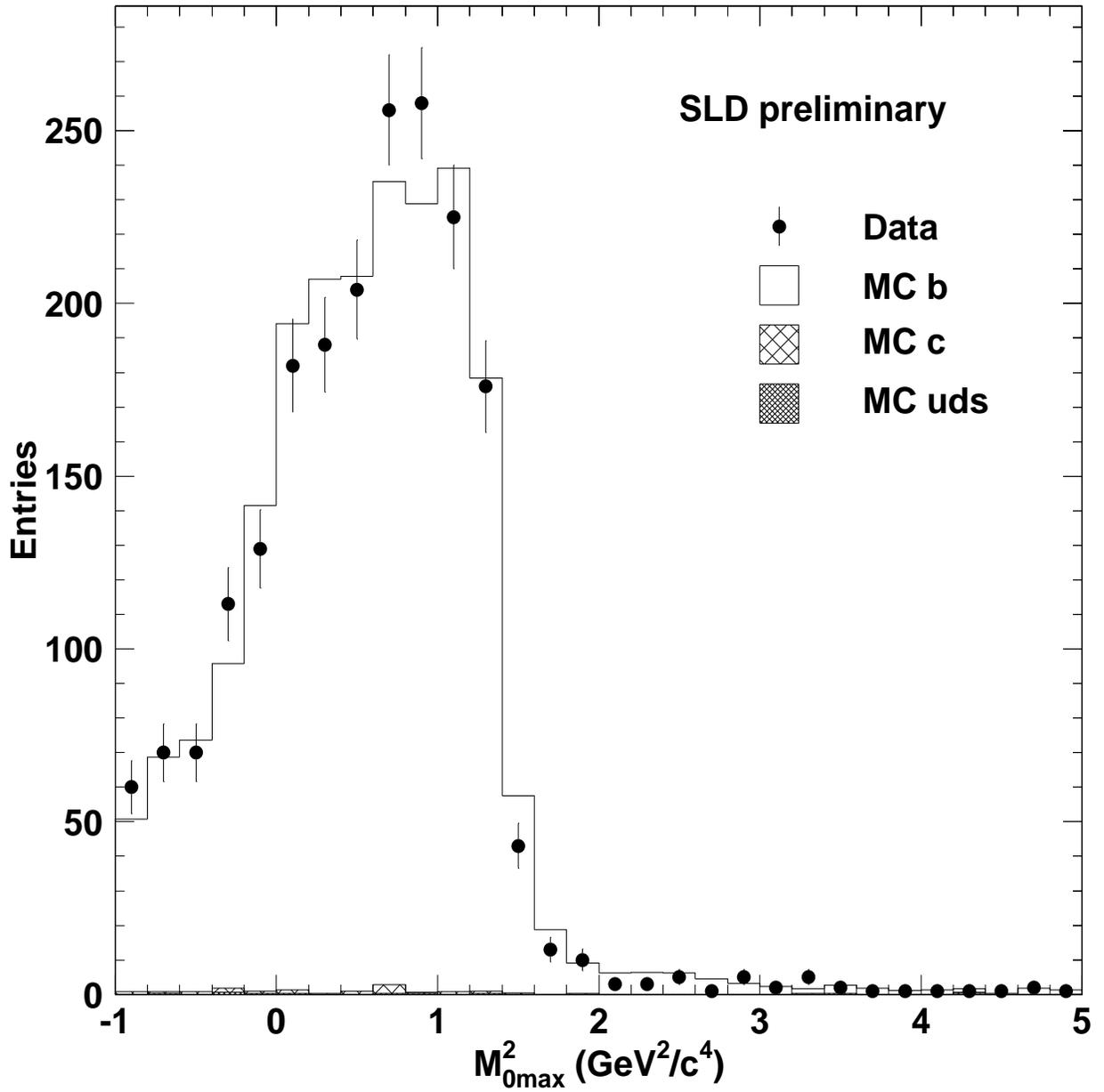}
}
\vskip -.2 cm
\caption[]{
\label{m0max_before}
Distribution of the reconstructed $M_{0max}^{2}$ for the
final selected $B$ sample (see text).
Also shown is the prediction of the Monte Carlo simulation.
}
\end{figure}

Only about 0.7\% of the selected vertices are from light-flavor events, 
but they are concentrated in the lowest energy bin.  To further remove 
this background, a vertex is required to contain at least 3 quality 
tracks with a normalized impact 
parameter greater than 2.  This eliminates almost all of the uds-event
background and cuts the charm background 
by about 20\% overall and 43\% in the few lowest energy bins.
This cut helps to reduce the dependence of the reconstructed $B$ hadron 
energy distribution on the light flavor simulation 
in the low energy region, which is a key step towards finding the correct 
shape of the $B$ hadron energy distribution at low energies.
Figure~\ref{m0max_before} shows the distribution of $M_{0max}^2$ after all 
these cuts; the data and Monte Carlo simulation are in good agreement.

\begin{figure}[ht]	
\centerline{\epsfxsize 1.1 truein 
\epsfysize6.5 in
\epsfxsize6.5 in
\leavevmode
\epsfbox{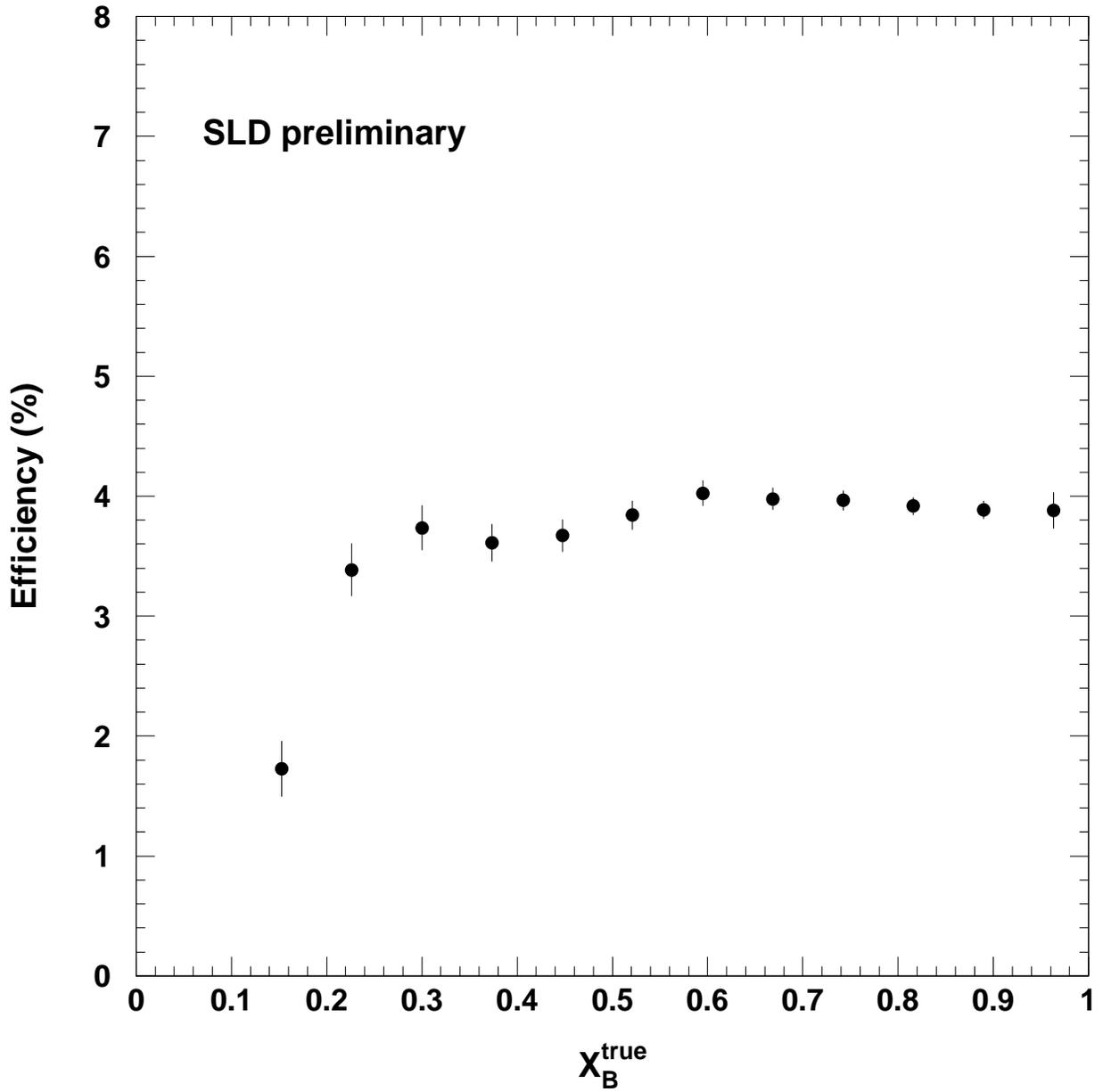}
}
\vskip -.2 cm
\caption[]{
\label{efficiency}
The Monte Carlo simulated efficiency for selecting 
 $B$ hadron decay vertices as a function of the 
 true scaled $B$ hadron energy, $x^{true}=E_{B}^{true}/E_{beam}$.
 The nearly energy-independent efficiency (except at very low $B$ energy)
improves the sensitivity
of the measured $x_B^{rec}$ distribution to the true underlying $B$ 
energy distribution.

}
\end{figure}

A total of 1920 vertices in the data for 1996-97 satisfy all these selection 
cuts.  
The overall efficiency for selecting $B$ hadrons is 3.9\% and the estimated 
$B$ hadron purity is 99.5\%.  
The efficiency as a function of $x_{B}^{true}$ is 
shown in Figure~\ref{efficiency}.  The dependence is rather weak
except for the lowest $x_B$ region; the efficiency is substantial, 
about 1.7\% even just above the kinematic threshold for $B$ energy.   

\begin{figure}[ht]	
\centerline{\epsfxsize 1.1 truein 
\epsfysize6.5 in
\epsfxsize6.5 in
\leavevmode
\epsfbox{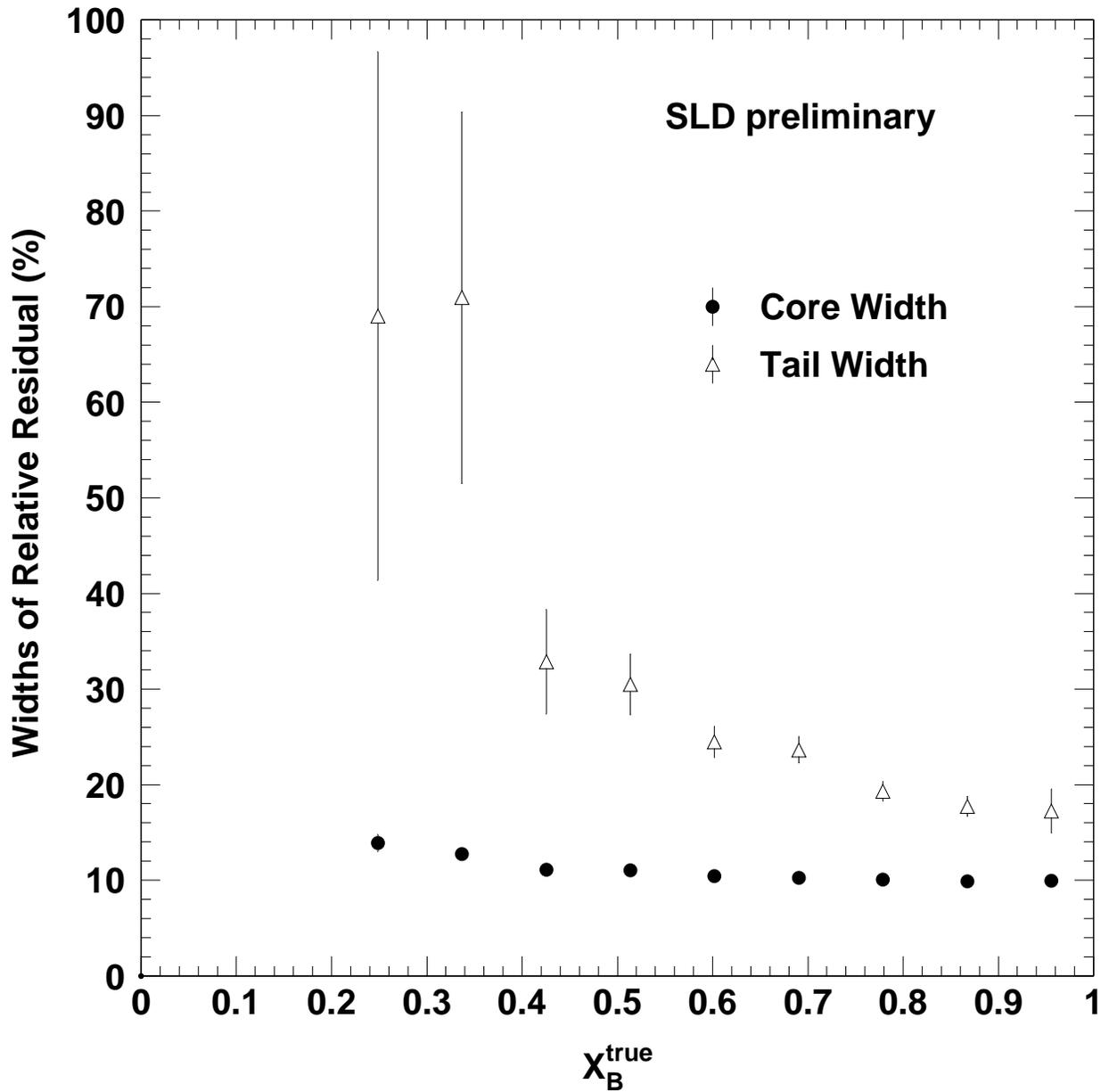}
}
\vskip -.2 cm
\caption[]{
\label{sigmavsx}
The fitted core and tail widths of the $B$ energy resolution as 
a function of the true scaled $B$ hadron energy. 
The ratio of the amplitude of the inner Gaussian (core) to that of 
the outer Gaussian (tail) is 83:17.  The dependence of the 
core resolution on the true $B$ energy is small.  The very 
good resolution for low energy $B$ hadrons improves the sensitivity
of the measured $x_B^{rec}$ distribution to the true underlying $B$ energy
distribution.
}
\end{figure}

We examine the $B$-energy resolution of this technique.
The distribution of the normalized difference 
between the true and reconstructed $B$ hadron energies, 
$(x_{B}^{rec}-x_{B}^{true})/x_{B}^{true}$, for Monte Carlo events,  
is fitted by a double Gaussian, resulting in a core width 
(the width of the narrower Gaussian) of 10.4\% and a tail width 
(the width of the wider Gaussian) of 23.6\% with a core fraction of 83\%.
Figure~\ref{sigmavsx} shows the core and tail widths as a function 
of $x_{B}^{true}$.  In order to compare the widths from different $x_B$ bins, 
we fix the ratio between core and tail fractions to that obtained in the 
overall fit above.  The $x_B$-dependence of the resolution is weak, 
indicating that the absolute resolution on $x_B$, $x_B^{rec}-x_B^{true}$,
is very good at low $B$ energy, which is an advantage of this energy 
reconstruction technique.

\begin{figure}[ht]	
\centerline{\epsfxsize 1.1 truein 
\epsfysize6.5 in
\epsfxsize6.5 in
\leavevmode
\epsfbox{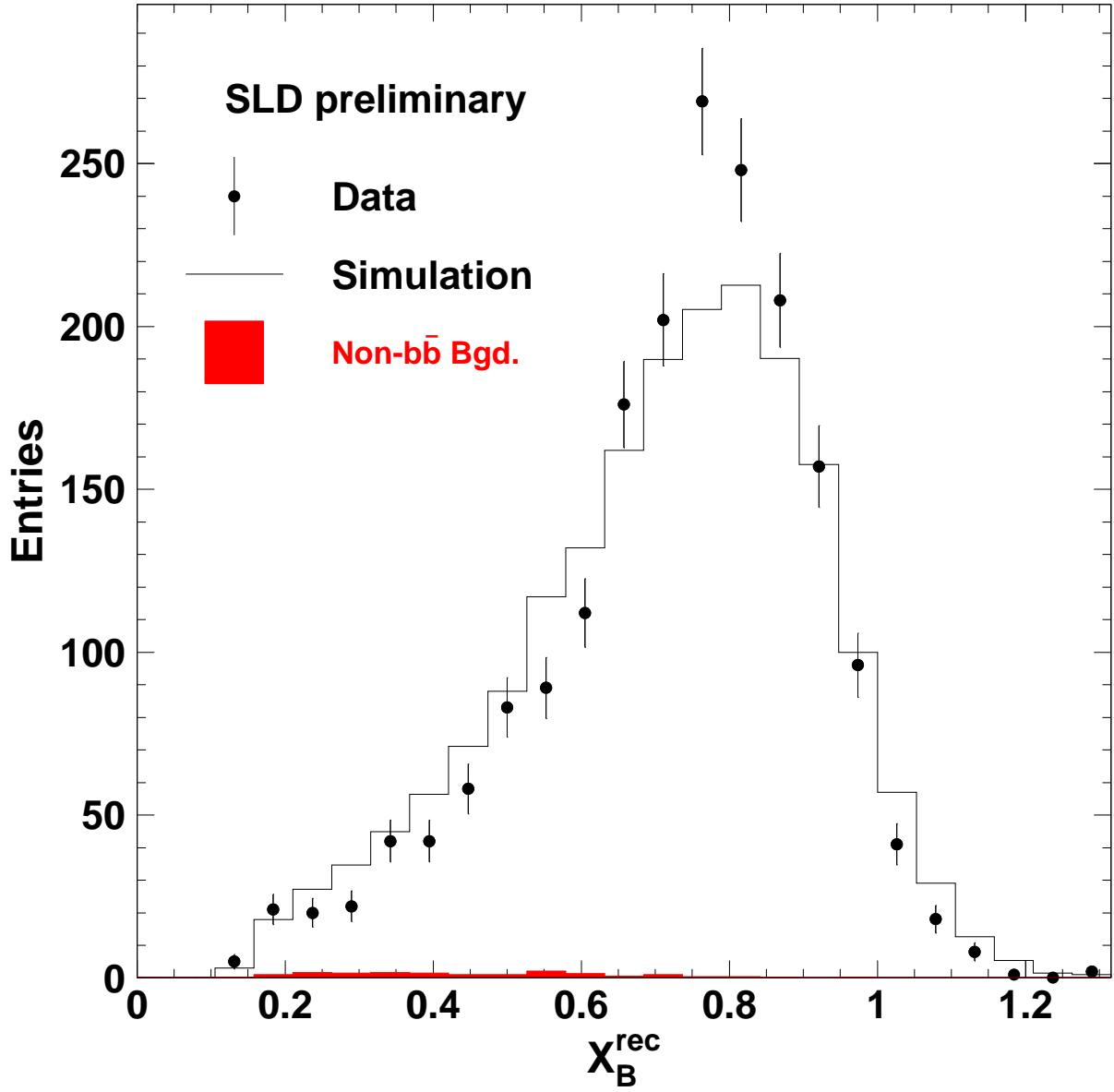}
}
\vskip -.2 cm
\caption[]{
\label{xbrec}
Distribution of the reconstructed scaled $B$ hadron energy for 1996-97 data 
(points) and the default Monte Carlo simulation (histogram).  The solid 
histogram shows the non-$b\bar{b}$ background.
}
\end{figure}

Figure~\ref{xbrec} shows the distribution of the reconstructed scaled 
$B$ hadron energy for the data, $D^{data}(x_{B}^{rec})$, and for the 
Monte Carlo simulation, $D^{MC}(x_{B}^{rec})$.  
The small non-\bb background, the high $B$ selection efficiency over 
the full kinematic coverage, and the good energy resolution
combine to give a much improved sensitivity of the data to the underlying 
true {\em shape} of the $B$ energy distribution (see next section).

The event generator used in our simulation is based on a perturbative QCD 
`parton shower' for production of quarks and gluons, together with the 
phenomenological Peterson function~\cite{pete} 
(Table~\ref{table:fragmodels}) to account for the fragmentation of $b$ and $c$ 
quarks into $B$ and $D$ hadrons, respectively,  
within the iterative Lund string hadronisation mechanism~\cite{jetset}; 
this simulation yields a `generator-level'
primary $B$ hadron energy distribution with 
$<x_{B}>$ = 0.693\footnote{We used a value of the Peterson function 
parameter $\epsilon_b$ = 0.006~\cite{sldrb}.}.
It is apparent that this simulation does not reproduce the data well 
(Figure~\ref{xbrec}); the $\chi^2$ for the comparison is 62 
for 16 bins\footnote{We exclude several bins with very few events in
the comparison. For details see Section~\ref{subsec:model} for details.}.

The distribution of the non-$b\bar{b}$ background, 
$S(x_{B}^{rec})$, is also shown in Figure~\ref{xbrec}.   
The background is subtracted bin-by-bin from the $D^{data}(x_B^{rec})$ 
before we proceed to test various fragmentation models.

\section{The Shape of the $B$ Hadron Energy Distribution}
\label{sec:shape}

Given the raw reconstructed $B$ energy distribution in the 
data shown in Figure~\ref{xbrec}, there are several ways of estimating the 
true underlying $B$ energy distribution.  Here we take
two approaches, each described in a subsection.

In the first part, we test several $b$ fragmentation models, 
$f(z,\beta)$ embedded within Monte Carlo generators, where $z$ is an
internal, experimentally inaccessible variable, corresponding roughly 
to the fraction of the momentum of the fragmentating $b$ quark carried by 
the resulting $B$ hadron, and $\beta$ is the set of parameters associated 
with the model in question.  
In the second part, we test several functional forms for the distribution
of $x_B$ itself, $f(x_B,\lambda)$, where $\lambda$ represents the set of 
parameters associated with each functional form.   

\subsection{Tests of $b$ Quark Fragmentation Models $f(z,\beta)$}
\label{subsec:model}

We first consider testing models of $b$ quark fragmentation.
Since the fragmentation functions for various models are 
usually functions of an experimentally inaccessible variable 
(e.g. $z=(E+p_{\|})_{H}/(E+p_{\|})_Q$ or $z = {p_{\|}}_H / {p_{\|}}_Q$ ), 
it is necessary to use a Monte Carlo generator 
to generate events according to a given input heavy
quark fragmentation function $f(z,\beta)$, where $\beta$ represents the
set of parameters.

We consider the phenonmenological models of 
the Lund group~\cite{lund}, Bowler~\cite{bowler},
Peterson \etal~\cite{pete} and Kartvelishvili \etal~\cite{kart}.  
We also consider the perturbative QCD calculations of Braaten 
\etal (BCFY)~\cite{bcfy}, and of Collins and Spiller (CS)~\cite{collins}. 
We use the JETSET~\cite{jetset} parton shower Monte Carlo and 
each fragmentation model in question to generate the simulated events 
without detector simulation.  
Table~\ref{table:fragmodels} contains a list of the models.
In addition, we test the UCLA~\cite{ucla} fragmentation model which
with fixed parameters.  For $b$ fragmentation, we also test
the HERWIG~\cite{herwig} event-generator using both possible values 
of the parameter $cldir=0$ and $1$.

\begin{table}[htb]
\begin{center}
\begin{tabular}{|l|c|c|}
\hline
Model  &  $f(z,\beta)$    &   Reference \\
\hline
BCFY &  $\frac{\textstyle z(1-z)^{2}}{\textstyle [1-(1-r)z]^{6}}[3+{\sum_{i=1}^{4} (-z)^{i}f_{i}(r)}]$  & \cite{bcfy} \\
Bowler  & $\frac{\textstyle 1}
{\textstyle z^{(1+r_{b}bm_{\perp}^{2})}}(1-z)^{a}exp(-bm_{\perp}^{2}/z)$
      & \cite{bowler} \\
CS &$ (\frac{\textstyle 1-z}{\textstyle z}+\frac{\textstyle (2-z)\epsilon_{b}}{\textstyle 1-z})
(1+z^{2})(1-\frac{\textstyle 1}{\textstyle z}-\frac{\textstyle \epsilon_{b}}{\textstyle 1-z})^{-2}$ & \cite{collins} \\
Kart.  & $z^{\alpha_{b}}(1-z)$ & \cite{kart} \\
Lund  & $\frac{\textstyle 1}{\textstyle z}(1-z)^{a}exp(-bm_{\perp}^{2}/z)$
      & \cite{lund} \\
Peterson & $\frac{\textstyle 1}{\textstyle z}(1-\frac{\textstyle 1}{\textstyle z}-\frac{\textstyle \epsilon_{b}}{\textstyle 1-\textstyle z})^{-2}$   & \cite{pete} \\
\hline
\end{tabular}
\caption{\label{table:fragmodels} 
$b$ quark fragmentation models used in comparison with the data.  
For the BCFY model, $f_{1}(r)~=~3(3-4r)$, 
$f_{2}(r)~=~12-23r+26r^{2}$, $f_{3}(r)~=~(1-r)(9-11r+12r^{2})$, and 
$f_{4}(r)~=~3(1-r)^{2}(1-r+r^{2})$. 
}
\end{center}
\end{table}

In order to make a consistent comparison of each model 
with the data we adopt the following procedure.  For each model, 
starting values of the arbitrary parameters, $\beta$, are assigned 
and the corresponding fragmentation function $f(z,\beta)$ is used
along with the JETSET Monte Carlo to produce the corresponding 
scaled primary $B$ hadron energy distribution, $D^{MC}(x_{B}^{true})$ 
in the MC-generated \bb event sample, {\it before} simulation of the 
detector.  Then each simulated $B$ hadron is weighted according to its 
true $B$ hadron energy, $x_B^{true}$; the weight is determined by the 
ratio of the generated $B$ hadron energy distribution, 
$D^{MC}(x_{B}^{true})$, to that of our default simulation
$D^{default}(x_{B}^{true})$.
After simulation of the detector, application of the analysis cuts and 
background subtraction, the resulting weighted distribution of reconstructed 
$B$ hadron energies, $D^{MC}(x_{B}^{rec})$, is then compared with the 
background-subtracted data distribution and the $\chi^2$ value, defined as
\begin{equation}
\chi^2 = \sum_{i=1}^{N} \left( \frac{\textstyle N_{i}^{data} - r N_{i}^{MC} }
{\textstyle \sigma_{i}} \right)^{2}
\label{eqn:chisq}
\end{equation}
is calculated, where $N$ is the number of bins to be used in the comparison, 
$N_{i}^{data}$ is the number of entries
in bin $i$ in the data distribution, and $N_{i}^{MC}$ is the number of entries 
in bin $i$ in the simulated distribution\footnote{$r$ is the factor by which 
the total number of entries 
in the simulated distribution is scaled to the number of entries in 
the data distribution; $r$ $\simeq$ 1/12.}.
$\sigma_{i}$ is the statistical error on the deviation of the 
observed number of entries for the data from the expected number of 
entries in bin $i$, which can be expressed as
\begin{equation}
\sigma_i^2 = \left( \sqrt{rN_i^{MC}} \right)^2 + 
\left( r\sqrt{N_i^{MC}} \right)^2,
\label{eqn:error}
\end{equation}
where $ \left( \sqrt{rN_{i}^{MC}} \right)^2$ is the expected statistical 
variance on the observed data number of entries in bin $i$, 
assuming the model being tested is correct, and 
$ \left( r\sqrt{N_{i}^{MC}} \right)^2 $ is the statistical variance on 
the expected number of entries in bin $i$.  Since the $\chi^{2}$-test is 
not a statistically effective test for bins with a very small number of 
entries, the third, the fourth, and the last three bins in Figure~\ref{xbrec} 
are excluded from the comparison.

\begin{figure}[ht]	
\centerline{\epsfxsize 1.1 truein 
\epsfysize6.5 in
\epsfxsize6.5 in
\leavevmode
\epsfbox{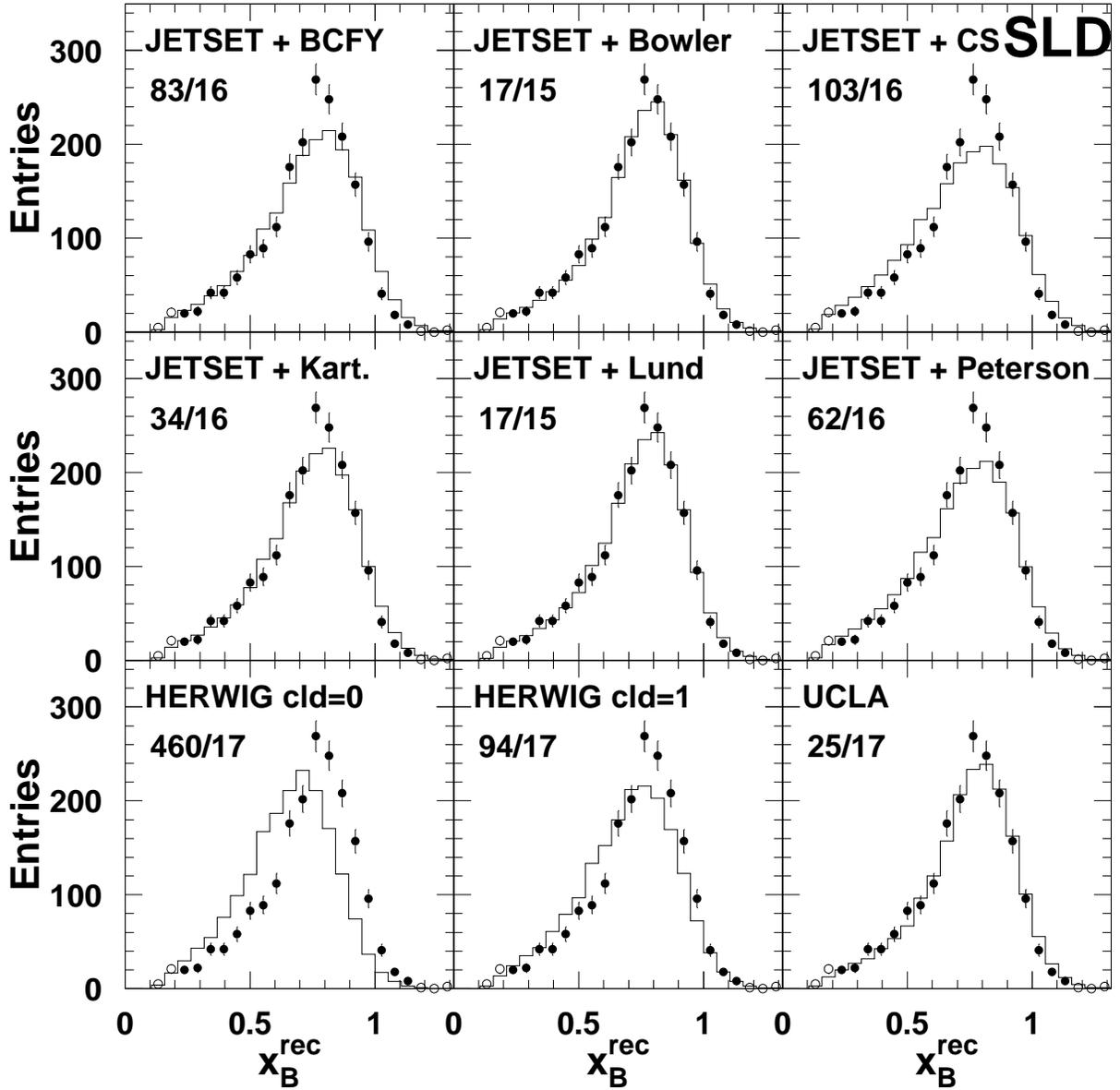}
}
\vskip -.2 cm
\caption[]{
\label{fig:fragmodel}
Each figure shows the background-subtracted distribution of 
reconstructed $B$ hadron energy for the data (points) and for the 
Monte Carlo (histogram) based on the respective optimised input 
fragmentation function within the JETSET parton shower simulation,  
as well as based on the HERWIG ($cld=0$ and $cld=1$) and the UCLA 
fragmentation models.  The $\chi^2$ and the number of degrees 
of freedom are indicated.
}
\end{figure}

We vary the values of the set of parameters $\beta$ and repeat the
above procedure.  The minimum $\chi^2$ is found by scanning through 
the input parameter space, yielding
a set of parameters which give an optimal description of the reconstructed
data by the fragmentation model in question.  
Each of the nine plots in Figure~\ref{fig:fragmodel} 
shows the background-subtracted distribution of 
reconstructed $B$ hadron energy for the data (points) and 
the respective $B$ energy distribution (histogram) resulting 
{\em either} from the optimised input fragmentation function $f(z)$ embedded
within the JETSET parton shower simulation, {\em or} from the predictions 
of the HERWIG event-generator and the UCLA fragmentation model.
Data points excluded from the fit are 
represented in Figure~\ref{fig:fragmodel} by open circles.
Table~\ref{table:modelresult} lists the results of the comparisons.
%
%


\begin{table}[htb]
\begin{center}
\begin{tabular}{|l|c|c|c|}
\hline
Model &  $\chi^{2}/dof$    &   Parameters &  $\langle x_{B} \rangle$\\
\hline
JETSET + BCFY
     &  83/16   & $r=0.085$ & 0.694 \\
JETSET + Bowler* &  17/15   & $a=1.5, b=1.5, (r_b=1)$          & 0.714 \\
JETSET + Collins and Spiller &  103/16   & $\epsilon_b=0.003$ & 0.691 \\
JETSET + Kartvelishvili* {\em et al.}  
      &  34/16   & $\alpha_b = 10.4$  &  0.711  \\
JETSET + Lund*   &  17/15   & $a=2.0, b=0.5$          & 0.712 \\
JETSET + Peterson {\em et al.}  & 62/16 & $\epsilon_{b}=0.006$  &  0.697 \\
HERWIG cldir=0  & 460/17 &   $-$   &  0.632 \\
HERWIG cldir=1  &  94/17 &   $-$    &  0.676 \\
UCLA* &  25/17 &   $-$    &  0.718 \\
\hline
\end{tabular}
\caption{\label{table:modelresult} 
Results of fragmentation model tests for JETSET + fragmentation models, 
the HERWIG model and the UCLA model.  Minimum $\chi^{2}$, 
number of degrees of freedom, coresponding parameter
values, and the mean value of the corresponding $B$ energy distribution 
are listed. * indicates used to correct the data in Section~\ref{sec:correct}.
}
\end{center}
\end{table}

We conclude that with our resolution and our current data sample, we
are able to distinguish between several fragmentation models.
Within the context of the JETSET Monte Carlo,  the Lund and Bowler 
models are consistent with the data with $\chi^2$ probability of
32\% for each, the Kartvelishvili model is marginally consistent with
the data, while the Peterson, the BCFY and the CS models are 
found to be inconsistent with the data.
The UCLA model is consistent with the data to a level of 10\% 
$\chi^2$ probability.  The HERWIG model with $cldir=0$ is confirmed to 
be much too soft.  Using $cldir=1$ results in a substantial improvement, 
but it is still inconsistent with the data. 

\subsection{Tests of Functional Forms $f(x_B,\lambda)$}
\label{subsec:form}
We then consider the more general question of what functional forms 
of the $B$ energy 
distribution, $f(x_B,\lambda)$, can be used as estimates of the true 
underlying $B$ energy distribution.
In particular, we would like to test a wide variety of functional forms
and ask how many different forms are consistent with the data.  Each 
consistent functional form will add to the list of our estimates of the
true underlying $B$ energy distribution.

%
For convenience we consider the functional forms of 
the BCFY, Collins and Spiller, Kartvelishvili, Lund, and Peterson 
groups in the variable $x_B$.
In addition we consider \adhoc generalisations of the Peterson function (F),
an 8th-order polynomial and a `power' function.  These functions are 
listed in Table~\ref{table:functionalform}. 
Each function vanishes at $x_{B}=0$ and $x_{B}=1$.

\begin{figure}[ht]	
\centerline{\epsfxsize 1.1 truein 
\epsfysize6.5 in
\epsfxsize6.5 in
\leavevmode
\epsfbox{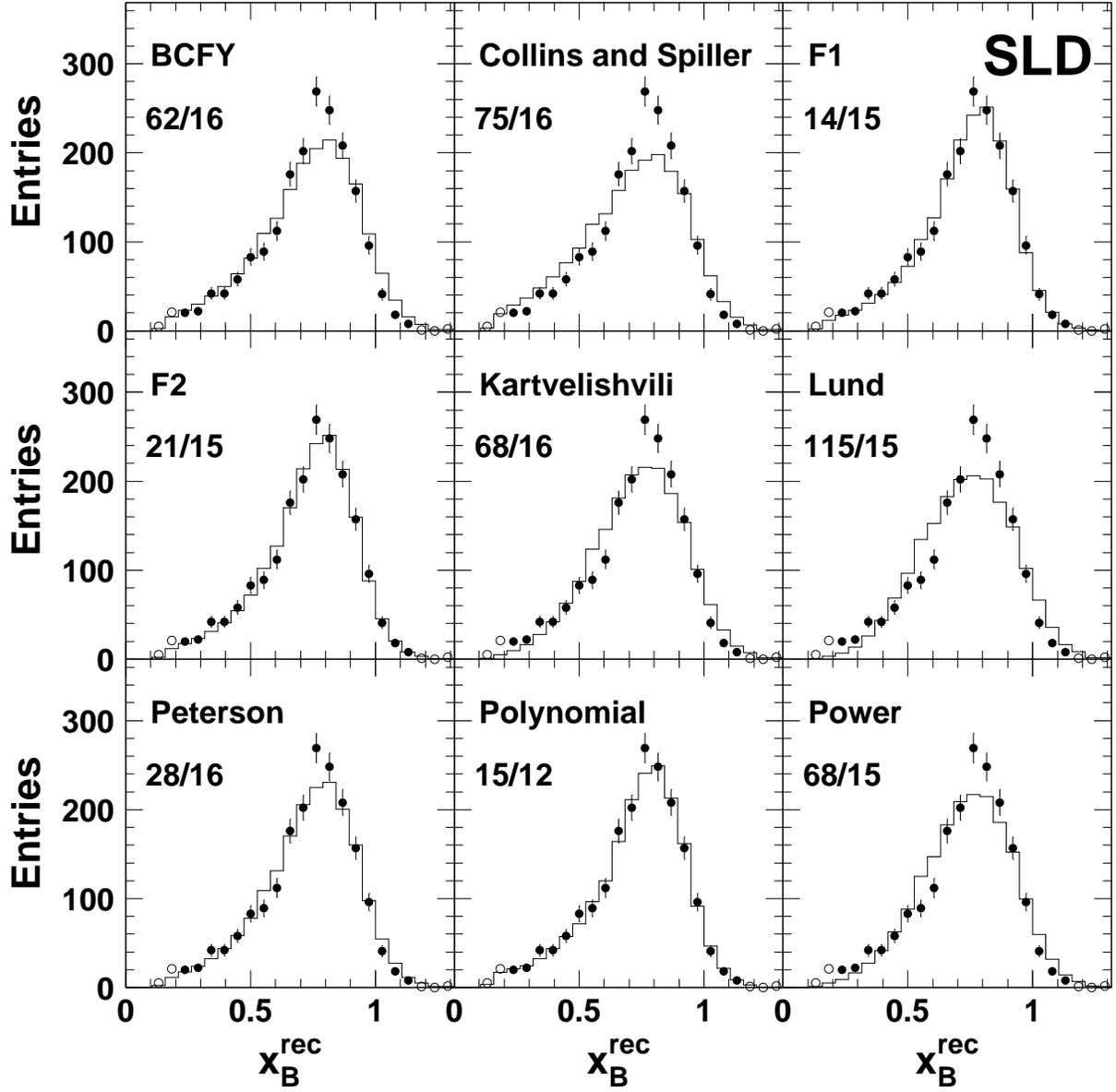}
\vspace{0.1cm}
}
\vskip -.2 cm
\caption[]{
\label{fig:form}
Each figure shows the background-subtracted distribution of 
reconstructed $B$ hadron energy for the data (points) and for the weighted 
simulation (histograms) based on the respective optimised input 
functional form for the {\em true} $B$ energy distribution.  
The $\chi^2$ and the number of degrees of freedom are indicated.
}
\end{figure}

\begin{table}[htb]
\begin{center}
\begin{tabular}{|l|c|c|} 
\hline
Function  &  $f(x_B,\lambda)$    &   Reference \\ 
\hline
F & $\frac{\textstyle (1+b(1-x_B))}{\textstyle x_B}(1-\frac{\textstyle c}{\textstyle x_B}-\frac{\textstyle d}{\textstyle 1-x_B})^{-2}$ & \cite{aleph95} \\

8th-order Polynomial & $x_B(1-x_B)(x_B-x_B^0)(1+{\sum_{i=1}^{5} p_{i}x_B^{i}})$  &   (see text)  \\
Power  & $x_B^{\alpha}(1-x_B)^{\beta}$  &  (see text) \\
\hline
\end{tabular}
\caption{\label{table:functionalform}
\small
$B$ energy functional forms used in comparison with the data. 
A polynomial function and a power function are included 
(see text for discussion).  $x_B^0$ is the low kinematic threshold for
$B$ energy.  For BCFY, CS, Kartvelishvili, Lund, Peterson functional
forms, see Table~\ref{table:fragmodels}.
}
\vspace{-0.5cm}
\end{center}
\end{table}
\begin{table}[htb]
\begin{center}
\begin{tabular}{|l|c|c|c|}
\hline
Function &  $\chi^{2}/dof$    &   Parameters &  $\langle x_{B} \rangle$\\
\hline
F1* & 14/15  & $c=0.838\pm0.018$ &    0.714$\pm$0.005 \\     
      &            & $d=0.022\pm0.002$           &     \\
F2* & 21/15  & $c=0.896\pm0.033$ & 0.717$\pm$0.005 \\ 
      &            & $d=0.040\pm0.003$           &     \\
BCFY  &  62/16   & $r=0.240\pm0.009$ & 0.709$\pm$0.005 \\
Collins and Spiller  
     &  75/16   & $\epsilon_{b}=0.043\pm0.005$ & 0.711$\pm$0.005 \\
Kartvelishvili {\em et al.}  
      &  68/16   & $\alpha_{b}=4.16\pm0.11$  &  0.721$\pm$0.004   \\
Lund  &  115/15  & $a=2.30\pm0.12$ & 0.721$\pm$0.005 \\
      &            & $bm_{\perp}^{2}=0.50\pm0.07$ &     \\
Peterson {\em et al.}*  & 28/16 & 
$\epsilon_{b}=0.036\pm0.002$ & 0.713$\pm$0.005 \\
Polynomial*
      & 15/12    &  $p_{1}=-10.76\pm0.16$             &  0.709$\pm$0.005   \\
      &            &  $p_{2}=45.74\pm0.28$    &   \\
            &          & $p_{3}=-93.60\pm0.34$  &   \\
            &          & $p_{4}=92.01\pm0.37$   &   \\
            &          & $p_{5}=-34.53\pm0.27$   &   \\
Power 
      &  68/15   & $\alpha=4.27\pm0.25$   & 0.720$\pm$0.005  \\
      &            & $\beta=1.05\pm0.10$  &                  \\
\hline
\end{tabular}
\caption{\label{table:formresult} 
Results of the $\chi^{2}$ fit of fragmentation functions to the reconstructed
$B$ hadron energy distribution after background subtraction.  The minimum 
$\chi^{2}$ value, the number of degrees of freedom, the coresponding 
parameter values, and the mean value of the corresponding $B$ energy 
distribution are listed.  Errors are statistical only. 
* indicates used to correct the data in Section~\ref{sec:correct}.
}
\end{center}
\end{table}
For each functional form, 
a testing procedure similar to that described in 
subsection~\ref{subsec:model} is applied. The optimised fitting 
parameters $\lambda$ 
and the minimum $\chi^2$ values are listed 
in Table~\ref{table:formresult}.  The corresponding $D^{MC}(x_{B}^{rec})$
are compared with the data in Figure~\ref{fig:form}.

Two sets of optimised parameters are found for the generalised
Peterson function F to describe the data.  
`F1', obtained by setting the parameter $b$ (shown in 
Table~\ref{table:functionalform}) to infinity, 
behaves like $x_B$ as $x_B$ \ra 0 and $(1-x_B)^3$ as $x_B$ \ra 1 and yields 
the best $\chi^2$ probability of 53\%; 
`F2', obtained by setting $b$ to zero, has a $\chi^2$ probability of 13\%.  
A constrained polynomial of at least 8th-order is needed to obtain
a $\chi^{2}$ probability greater than 0.1\%.
The Peterson functional form marginally reproduces the data with a 
$\chi^2$ probability of about 3\%.
The remaining functional forms are found to be inconsistent with our data.
The widths of the BCFY and CS functions are too large to 
describe the data;  Kartvelishvili, Lund and the `power'
functional form vanish too fast as $x_B$ approaches zero.  
We conclude that, within our resolution and with our 
current data sample, we are able to distinguish between some of these 
functional forms.  But most importantly, consistent functional forms 
will help us evaluate the uncertainty on 
the true $B$ energy distribution.

\section{Correction of the $B$ Energy Distribution}
\label{sec:correct}

In order to compare our results with those from other experiments and
potential future theoretical predictions it is
necessary to correct the reconstructed scaled $B$ hadron energy distribution 
$D^{data}(x_{B}^{rec})$ for the 
effects of non-$B$ backgrounds, detector acceptance, event selection and
analysis bias, and initial-state radiation, as well as for bin-to-bin
migration effects caused by the finite resolution of the detector and the
analysis technique. 
%
%
Due to the known rapid variation of the yet-unknown true $B$ energy
distribution at large $x_B$, {\em any} correction procedure will 
necessarily be more or less model-dependent.  
We choose a method that explicitly 
evaluates this model-dependence and gives a very good estimate of the 
true energy distribution using all of the above models or functional 
forms that are at least marginally consistent with the data.

We apply a $25\times25$ matrix unfolding procedure 
to $D^{data}(x_{B}^{rec})$ to obtain an estimate of the true distribution 
$D^{data}(x_{B}^{true})$:
\vspace{-0.12cm}
\begin{eqnarray}
D^{data}(x_{B}^{true})\quad=\quad \epsilon^{-1}(x_{B}^{true}) \cdot 
E(x_{B}^{true},x_{B}^{rec}) \cdot (D^{data}(x_{B}^{rec})
-S(x_{B}^{rec})) 
\label{eqn:unfold}
\vspace{-0.5cm}
\end{eqnarray}
where $S$ is a vector representing the background contribution, $E$ is a
matrix to correct for bin-to-bin migrations, and $\epsilon$ is
a vector representing the efficiency for selecting true $B$ hadron
decays for the analysis. 
The matrices $S$, $E$ and $\epsilon$ are calculated from our MC
simulation; 
%
%
the matrix $E$ incorporates a
convolution of the input fragmentation function with the resolution of the
detector.  $E(i,j)$ is the number of vertices with $x_{B}^{true}$ in bin $i$ 
and $x_{B}^{rec}$ in bin $j$, normalized by the total number of vertices 
with $x_{B}^{rec}$ in bin $j$.  
%

\begin{figure}[ht]	
\centerline{\epsfxsize 1.1 truein 
\epsfysize6.5 in
\epsfxsize6.5 in
\leavevmode
\epsfbox{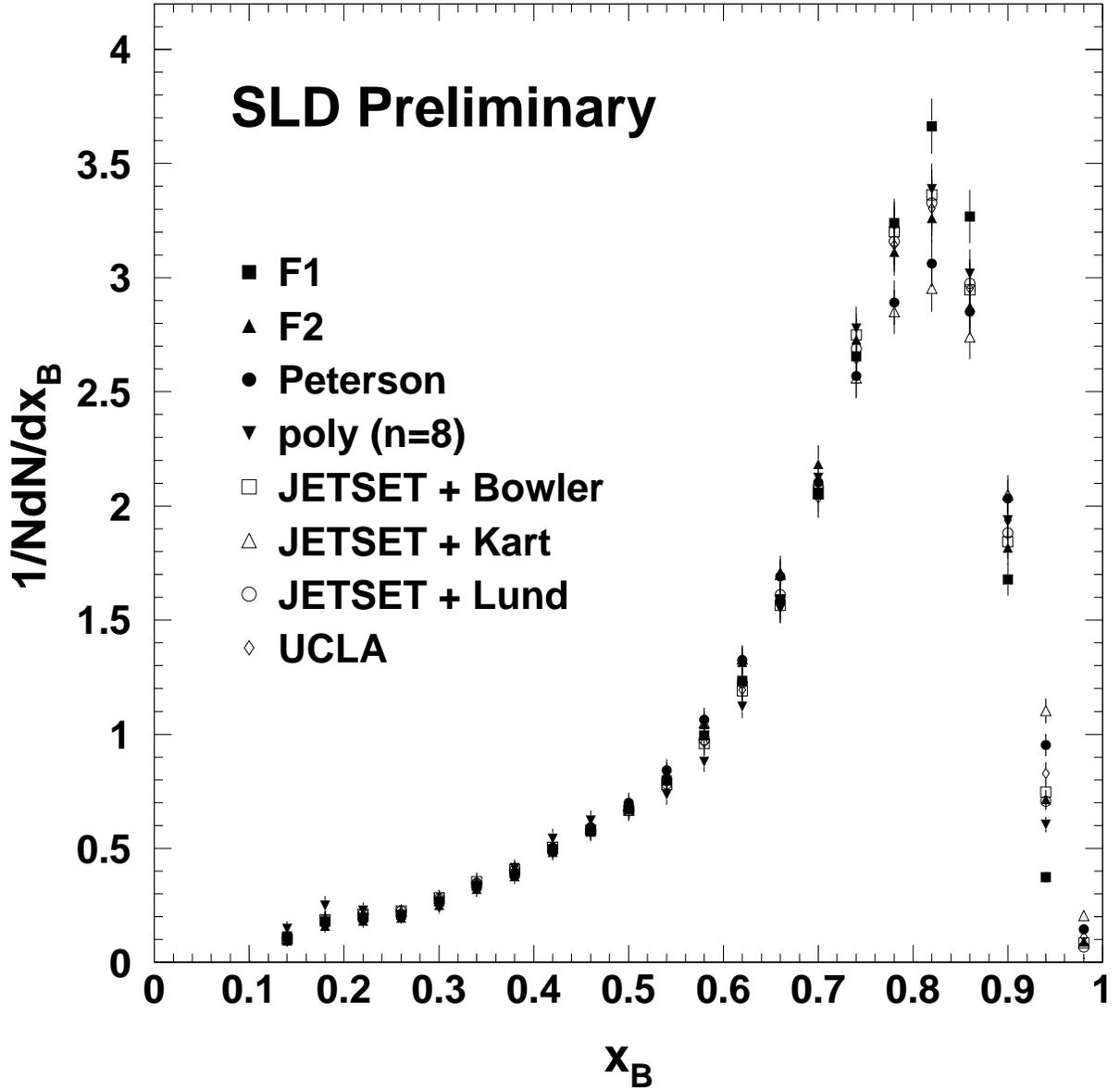}
}
\vskip -.2 cm
\caption[]{
\label{overlay}
The efficiency-resolution corrected distributions of scaled 
weakly-decaying $B$ hadron energies for {\bf Case 1)} fragmentation 
models of the Lund, the Bowler and the Karvelishvili 
within the JETSET parton shower Monte Carlo as well as for the UCLA 
fragmentation model; and for {\bf Case 2)} four functional forms: 
F1, F2, Peterson, and the constrained 8th-order polynomial. 
}
\end{figure}

We evaluate the matrix $E$ using the Monte Carlo simulation weighted 
according to an input generator-level {\em true} $B$ energy 
distribution found to be consistent with the data in 
Section~\ref{sec:shape}.  We have seen that several $B$ energy 
distributions can reproduce the data.
We consider in turn each of these eight consistent distributions, 
using the optimised parameters listed in Table~\ref{table:modelresult}
and~\ref{table:formresult}.
The matrix $E$ is then evaluated by examining 
the population migrations of true $B$ hadrons between bins 
of the input scaled $B$ energy, $x_{B}^{true}$, and 
the reconstructed scaled $B$ energy, $x_{B}^{rec}$. 
Using each $D^{MC}(x_{B}^{true})$, the data distribution 
$D^{data}(x_{B}^{rec})$ is then unfolded
according to Equation~(\ref{eqn:unfold}) to yield $D^{data}(x_{B}^{true})$, 
which is shown for each input fragmentation function in Figure~\ref{overlay}.

\begin{figure}[ht]	
\centerline{\epsfxsize 1.1 truein 
\epsfysize6.5 in
\epsfxsize6.5 in
\leavevmode
\epsfbox{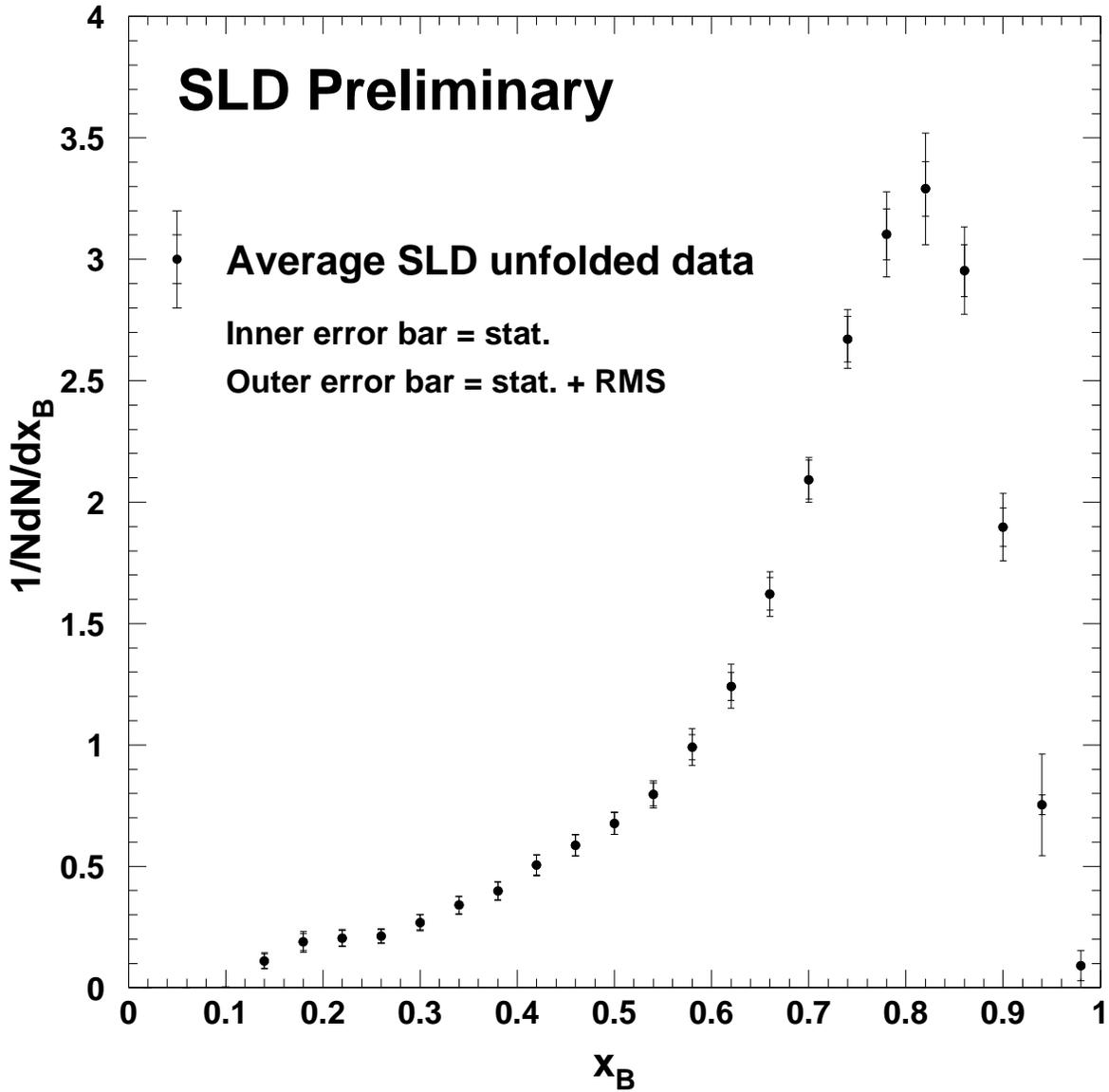}
}
\vskip -.2 cm
\caption[]{
\label{average}
Distribution of the final corrected scaled $B$ hadron energies.  The
central value is the bin-by-bin average of the eight consistent $B$
energy distributions.  In each bin the statistical error is indicated 
by the inner error bar, the sum in quadrature 
of statistical and unfolding errors from model dependence 
by the outer error bar.  Systematic errors are small compared with
the statistical and model dependence errors and are not included here.  
Note that the first two bins are below the kinematic limit for 
$x_{B}$ (no point shown).
}
\end{figure}

It can be seen that the shapes of $D^{data}(x_{B}^{true})$ differ 
systematically among the input $b$ quark fragmentation models 
and the assumed $B$ energy functional forms.
These differences are used to assign systematic errors.
Figure~\ref{average} shows the final corrected $x_{B}$ 
distribution $D(x_{B})$, which is the bin-by-bin average of 
the eight unfolded distributions,  
where the inner error bar represents the statistical error
and the outer error bar 
is the sum in quadrature of the r.m.s.\ of the eight unfolded distributions 
and the statistical error within each bin.  
Since two of the eight functions (the Kartvelishvili model and the 
Peterson functional form) are only in marginal agreement with the data,
and the 8th-order polynomial has a slightly unphysical behavior
near $x_B=1$, this r.m.s. may be considered to be a rather reasonable 
envelope within which the true $x_B$ 
distribution is most likely to vary.  The model dependence for this
analysis is significantly smaller than those of previous direct $B$ 
energy measurements, indicating
the enhanced sensitivity of our data to the underlying true energy 
distribution.    
  
\section{Systematic Errors}
\label{sec:sys}

We have considered sources of systematic uncertainty that potentially affect 
our measurement of the $B$ hadron energy distribution. 
These may be divided into uncertainties in 
modelling the detector and uncertainties on 
experimental measurements serving as
input parameters to the underlying physics modelling. 
For these studies our standard simulation, employing 
the Peterson fragmentation function, is used.

For each source of systematic error, the Monte Carlo distribution 
$D^{MC}(x_B^{true})$ is re-weighted and then the resulting 
Monte Carlo reconstructed distribution $D^{MC}(x_B^{rec})$ is 
compared with the data $D^{data}(x_B^{rec})$ 
by repeating the fitting and unfolding procedures described in Section 4
and 5.  
The differences in both the shape and the mean value of the $x_B^{true}$ 
distribution
relative to the standard procedure with nominal values of parameters 
are considered.  
Due to the strong dependence of our energy reconstruction technique 
on charged tracks, the dominant systematic error is due to the 
discrepancy in the charged track transverse momentum resolution between 
the Monte Carlo and the data.  
%
We evaluate this conservatively by taking the 
full difference between the nominal results and results using 
a resolution-corrected Monte Carlo event sample.
The difference between the measured and simulated charged track multiplicity
as a function of cos$\theta$ and momentum is attributed to an unsimulated 
tracking inefficiency correction.  We use a random track-tossing 
procedure to evaluate the difference in our results.  

\begin{table}[htb]
\begin{center}
\begin{tabular}{|l|c|r|}
\hline
Source & Variation & $\delta$ $\langle x_B \rangle$ \\
\hline
{\bf Monte Carlo statistics}  &   & {\bf 0.0011} \\
\hline
 Tracking efficiency correction &  on/off	&     0.0022\\     
 Track impact parameter &  on/off &  0.0012 \\
 Track polar angle  & 	2 mrad	&   $-$0.0009	\\ 
 Track $1/P_\perp$	&  0.0017  &  $-$0.0054 \\
 Hadronic event selection   &  standard &  0.0005 \\
\hline
{\bf Total Detector Systematics} & & {\bf 0.0061} \\
\hline
 $B^0$ mass effect  &  $\pm$2$\sigma$&  0.0001 \\
 $B$ lifetimes  & $\pm\sigma$&  0.0002 \\
 $B^+/B^0/B^0_s/\Lambda_b$ production & $\pm$2$\sigma$&  0.0010 \\
 $B$ decay fraction & $\pm$2$\sigma$ &  0.0006 \\
 $B$ decay $<n_{ch}>$& 5.3$\pm$0.3 &  0.0012 \\
 $D$ lifetimes & $\pm$$\sigma$ & 0.0002 \\
 $D$ decay $<n_{ch}>$& $\pm$$\sigma$  & 0.0005 \\
 $D \rightarrow K^0$ multiplicity & $\pm$$\sigma$ & 0.0013 \\
 $D \rightarrow$ no $\pi^0$ fraction & $\pm$$\sigma$ & 0.0005 \\
 $g \rightarrow b\bar{b}$ & (0.31$\pm$0.15)$\%$ & 0.0002 \\
 $g \rightarrow c\bar{c}$ & (2.4$\pm$1.2)$\%$ & 0.0008 \\
 $K^0$ production& 0.66$\pm$0.07 trks&0.0009 \\
 $\Lambda$ production& 0.12$\pm$0.01 trks&0.0002 \\
 $R_b$& 0.2170$\pm$0.0009&$<$0.0001 \\
 $R_c$& 0.1733$\pm$0.0048&$<$0.0001 \\
 Model dependence & & 0.0020 \\
\hline
{\bf Total Systematics}  & & {\bf 0.0068}\\
\hline
\end{tabular}
\caption{
\label{table:syst} 
Source and systematic errors.
}
\end{center}
\end{table}

\indent
A large number of measured quantities relating to the production and decay
of charm and bottom hadrons are used as input to our simulation. 
In \bb events we have considered the uncertainties on: 
the branching fraction for \z0 \ra \bb;
the rates of production of $B^{\pm}$, $B^0$ and $B^0_s$ mesons, 
and $B$ baryons;
the lifetimes of $B$ mesons and baryons;
and the average $B$ hadron decay charged multiplicity.
In \cc events we have considered the uncertainties on: 
the branching fraction for \z0 \ra \cc;
the charmed hadron lifetimes,
the charged multiplicity of charmed hadron decays,
the production of  $K^0$ from charmed hadron decays,
and the fraction of charmed hadron decays containing no $\pi^0$s.
We have also considered the rate of production of \ss in the jet fragmentation
process, and the production of secondary \bb and \cc from gluon splitting.
The world-average values~\cite{heavy,sldrb} of these quantities used in our
simulation, as well as the respective uncertainties, are listed in
Table~\ref{table:syst}.  Most of these variations have effect on
normalisation, but very little on the shape or the mean value.  In no
case do we find a variation that changes our conclusion about which
functions are consistent with the data.  Systematic errors of the mean value
are listed in Table~\ref{table:syst}.

The model-dependence of the unfolding procedure is estimated by considering
the envelope of the unfolded results illustrated in Figure~\ref{average}.
Since eight functions provide an acceptable $\chi^2$ probablity 
in fitting to the data, in each bin of $x_{B}$ we calculated the 
average value of these eight unfolded results as well as the r.m.s. 
deviation.  In each bin the average value is taken as our central
value and the r.m.s.\@ value is assigned as the unfolding uncertainty.

\indent
Other relevant systematic effects such as variation of 
the event selection cuts and the assumed $B$ hadron mass are also 
found to be very small.  As a cross-check, we vary the $M_{0max}$ 
cut (Equation~(\ref{eqn:m0maxcut})) in selecting the final $B$ sample 
within a large range and repeat the
analysis procedure.  In each case, conclusions about the shape of the $B$ 
energy distribution hold.  In each bin, all sources of systematic 
uncertainty are added in quadrature to obtain the total systematic error.


\section{Summary and Conclusions}

We have used the excellent tracking and vertexing capabilities of SLD 
to reconstruct the energies of $B$ hadrons in \ep \ra \z0 events over 
the full kinematic range by applying a new kinematic technique to 
an {\em inclusive} sample of topologically reconstructed $B$ hadron
decay vertices.  The overall $B$ selection efficiency of the
method is 3.9\%.
We estimate the resolution on the $B$ energy to be about 10.4\% for 
roughly 83\% of the reconstructed decays.  The energy resolution for
low energy $B$ hadrons is significantly better than previous measurements.

In order to get a good estimate of the model 
dependence of the unfolded distribution, 
the distribution of reconstructed scaled 
$B$ hadron energy, $D^{data}(x^{rec}_{B})$, is 
compared {\bf case 1)} with 
predictions of {\em either} perturbative QCD and phenomenological 
$b$ quark fragmentation models in the context of the JETSET 
parton shower Monte Carlo, 
{\em or} HERWIG and UCLA fragmentation models, and {\bf case 2)} 
with a set of functional forms for the $B$ energy distribution.  
In {\bf case 1)}, 
the Lund and the Bowler models are consistent with the data; 
the model of Kartvelishvili \etal is in marginal agreement with the data.  
The models based on the 
perturbative QCD calculations of Braaten \etal, and of Collins and Spiller, 
and the Peterson model are disfavored by the data.  Although both
versions of the HERWIG model are excluded by the data, the new 
version is very much improved.  
The UCLA model describes the data reasonably well.
In {\bf case 2)}, four functional forms, 
namely the two generalised Peterson functions F1 and F2, 
the Peterson function, and a constrained 8th-order polynomial 
are found to be consistent with the data. 

The raw $B$ energy distribution is then corrected 
for bin-to-bin migrations caused by the resolution of the method 
and for selection efficiency to derive the energy distribution 
of the weakly decaying $B$ hadrons produced in \z0 decays.  
Systematic uncertainties in the correction have been evaluated 
and are found to be significantly smaller than those of previous 
direct $B$ energy measurements.  The final corrected $x_{B}$ distribution 
$D^{data}(x_{B}^{true})$ is shown in Figure~\ref{average}.
The statistical and unfolding uncertainties are
indicated separately.

It is conventional to evaluate the mean of this $B$ energy 
distribution, $<x_{B}>$.
For each of the eight functions providing a reasonable description of
 the data (four from {\bf case 1)} and four from {\bf case 2)}), we 
evaluate $<x_{B}>$ from the distribution that corresponds to the 
optimised parameters; 
these are listed in Table~\ref{table:modelresult} and 
Table~\ref{table:formresult}.  We take the average 
of the eight values of $<x_{B}>$ as our central value, and 
define the model-dependent uncertainty to be the r.m.s.\@ deviation
within each bin.  
All detector and physics modeling systematic errors are included.  
We obtain 
\begin{eqnarray}
<x_{B}>\quad=\quad 0.714\pm 0.005 (stat.)\pm 0.007 (syst)\pm 0.002 (model),
\label{eqn:average}
\end{eqnarray}
It can be seen that $<x_{B}>$ is relatively insensitive to the variety of 
allowed forms of the shape of the fragmentation function $D(x_{B})$.

\section*{Acknowledgements}
We thank the personnel of the SLAC accelerator department and the
technical
staffs of our collaborating institutions for their outstanding efforts
on our behalf.

\vskip .5truecm
\small
\vbox{\footnotesize\renewcommand{\baselinestretch}{1}\noindent
$^*$Work supported by Department of Energy
  contracts:

  DE-FG02-91ER40676 (BU),
  DE-FG03-91ER40618 (UCSB),
  DE-FG03-92ER40689 (UCSC),

  DE-FG03-93ER40788 (CSU),
  DE-FG02-91ER40672 (Colorado),
  DE-FG02-91ER40677 (Illinois),

  DE-AC03-76SF00098 (LBL),
  DE-FG02-92ER40715 (Massachusetts),
  DE-FC02-94ER40818 (MIT),

  DE-FG03-96ER40969 (Oregon),
  DE-AC03-76SF00515 (SLAC),
  DE-FG05-91ER40627 (Tennessee),

  DE-FG02-95ER40896 (Wisconsin),
  DE-FG02-92ER40704 (Yale);

  National Science Foundation grants:

  PHY-91-13428 (UCSC),
  PHY-89-21320 (Columbia),
  PHY-92-04239 (Cincinnati),

  PHY-95-10439 (Rutgers),
  PHY-88-19316 (Vanderbilt),
  PHY-92-03212 (Washington);

  The UK Particle Physics and Astronomy Research Council
  (Brunel, Oxford and RAL);

  The Istituto Nazionale di Fisica Nucleare of Italy

  (Bologna, Ferrara, Frascati, Pisa, Padova, Perugia);

  The Japan-US Cooperative Research Project on High Energy Physics
  (Nagoya, Tohoku);

  The Korea Research Foundation (Soongsil, 1997).}


\vfill
\eject

\section*{$^{**}$List of Authors}
%
%
%
\begin{center}
\def\iADEL{$^{(1)}$}
\def\iAOMORI{$^{(2)}$}
\def\iBOLO{$^{(3)}$}
\def\iBRI{$^{(4)}$}
\def\iBRUN{$^{(5)}$}
\def\iBU{$^{(6)}$}
\def\iCINC{$^{(7)}$}
\def\iCOLO{$^{(8)}$}
\def\iCOLU{$^{(9)}$}
\def\iCSU{$^{(10)}$}
\def\iFERR{$^{(11)}$}
\def\iFRAS{$^{(12)}$}
\def\iILLI{$^{(13)}$}
\def\iJHU{$^{(14)}$}
\def\iLBL{$^{(15)}$}
\def\iLTU{$^{(16)}$}
\def\iMASS{$^{(17)}$}
\def\iMISSI{$^{(18)}$}
\def\iMIT{$^{(19)}$}
\def\iMOSCOW{$^{(20)}$}
\def\iNAGO{$^{(21)}$}
\def\iOREG{$^{(22)}$}
\def\iOXF{$^{(23)}$}
\def\iPADO{$^{(24)}$}
\def\iPERU{$^{(25)}$}
\def\iPISA{$^{(26)}$}
\def\iRAL{$^{(27)}$}
\def\iRUTG{$^{(28)}$}
\def\iSLAC{$^{(29)}$}
\def\iSOGA{$^{(30)}$}
\def\iSOONG{$^{(31)}$}
\def\iTENN{$^{(32)}$}
\def\iTOHO{$^{(33)}$}
\def\iUCSB{$^{(34)}$}
\def\iUCSC{$^{(35)}$}
\def\iUVIC{$^{(36)}$}
\def\iVAND{$^{(37)}$}
\def\iWASH{$^{(38)}$}
\def\iWISC{$^{(39)}$}
\def\iYALE{$^{(40)}$}

  \baselineskip=.75\baselineskip  
\mbox{Kenji  Abe\unskip,\iNAGO}
\mbox{Koya Abe\unskip,\iTOHO}
\mbox{T. Abe\unskip,\iSLAC}
\mbox{I.Adam\unskip,\iSLAC}
\mbox{T.  Akagi\unskip,\iSLAC}
\mbox{N. J. Allen\unskip,\iBRUN}
\mbox{W.W. Ash\unskip,\iSLAC}
\mbox{D. Aston\unskip,\iSLAC}
\mbox{K.G. Baird\unskip,\iMASS}
\mbox{C. Baltay\unskip,\iYALE}
\mbox{H.R. Band\unskip,\iWISC}
\mbox{M.B. Barakat\unskip,\iLTU}
\mbox{O. Bardon\unskip,\iMIT}
\mbox{T.L. Barklow\unskip,\iSLAC}
\mbox{G. L. Bashindzhagyan\unskip,\iMOSCOW}
\mbox{J.M. Bauer\unskip,\iMISSI}
\mbox{G. Bellodi\unskip,\iOXF}
\mbox{R. Ben-David\unskip,\iYALE}
\mbox{A.C. Benvenuti\unskip,\iBOLO}
\mbox{G.M. Bilei\unskip,\iPERU}
\mbox{D. Bisello\unskip,\iPADO}
\mbox{G. Blaylock\unskip,\iMASS}
\mbox{J.R. Bogart\unskip,\iSLAC}
\mbox{G.R. Bower\unskip,\iSLAC}
\mbox{J. E. Brau\unskip,\iOREG}
\mbox{M. Breidenbach\unskip,\iSLAC}
\mbox{W.M. Bugg\unskip,\iTENN}
\mbox{D. Burke\unskip,\iSLAC}
\mbox{T.H. Burnett\unskip,\iWASH}
\mbox{P.N. Burrows\unskip,\iOXF}
\mbox{A. Calcaterra\unskip,\iFRAS}
\mbox{D. Calloway\unskip,\iSLAC}
\mbox{B. Camanzi\unskip,\iFERR}
\mbox{M. Carpinelli\unskip,\iPISA}
\mbox{R. Cassell\unskip,\iSLAC}
\mbox{R. Castaldi\unskip,\iPISA}
\mbox{A. Castro\unskip,\iPADO}
\mbox{M. Cavalli-Sforza\unskip,\iUCSC}
\mbox{A. Chou\unskip,\iSLAC}
\mbox{E. Church\unskip,\iWASH}
\mbox{H.O. Cohn\unskip,\iTENN}
\mbox{J.A. Coller\unskip,\iBU}
\mbox{M.R. Convery\unskip,\iSLAC}
\mbox{V. Cook\unskip,\iWASH}
\mbox{R. Cotton\unskip,\iBRUN}
\mbox{R.F. Cowan\unskip,\iMIT}
\mbox{D.G. Coyne\unskip,\iUCSC}
\mbox{G. Crawford\unskip,\iSLAC}
\mbox{C.J.S. Damerell\unskip,\iRAL}
\mbox{M. N. Danielson\unskip,\iCOLO}
\mbox{M. Daoudi\unskip,\iSLAC}
\mbox{N. de Groot\unskip,\iBRI}
\mbox{R. Dell'Orso\unskip,\iPERU}
\mbox{P.J. Dervan\unskip,\iBRUN}
\mbox{R. de Sangro\unskip,\iFRAS}
\mbox{M. Dima\unskip,\iCSU}
\mbox{A. D'Oliveira\unskip,\iCINC}
\mbox{D.N. Dong\unskip,\iMIT}
\mbox{M. Doser\unskip,\iSLAC}
\mbox{R. Dubois\unskip,\iSLAC}
\mbox{B.I. Eisenstein\unskip,\iILLI}
\mbox{V. Eschenburg\unskip,\iMISSI}
\mbox{E. Etzion\unskip,\iWISC}
\mbox{S. Fahey\unskip,\iCOLO}
\mbox{D. Falciai\unskip,\iFRAS}
\mbox{C. Fan\unskip,\iCOLO}
\mbox{J.P. Fernandez\unskip,\iUCSC}
\mbox{M.J. Fero\unskip,\iMIT}
\mbox{K.Flood\unskip,\iMASS}
\mbox{R. Frey\unskip,\iOREG}
\mbox{J. Gifford\unskip,\iUVIC}
\mbox{T. Gillman\unskip,\iRAL}
\mbox{G. Gladding\unskip,\iILLI}
\mbox{S. Gonzalez\unskip,\iMIT}
\mbox{E. R. Goodman\unskip,\iCOLO}
\mbox{E.L. Hart\unskip,\iTENN}
\mbox{J.L. Harton\unskip,\iCSU}
\mbox{A. Hasan\unskip,\iBRUN}
\mbox{K. Hasuko\unskip,\iTOHO}
\mbox{S. J. Hedges\unskip,\iBU}
\mbox{S.S. Hertzbach\unskip,\iMASS}
\mbox{M.D. Hildreth\unskip,\iSLAC}
\mbox{J. Huber\unskip,\iOREG}
\mbox{M.E. Huffer\unskip,\iSLAC}
\mbox{E.W. Hughes\unskip,\iSLAC}
\mbox{X.Huynh\unskip,\iSLAC}
\mbox{H. Hwang\unskip,\iOREG}
\mbox{M. Iwasaki\unskip,\iOREG}
\mbox{D. J. Jackson\unskip,\iRAL}
\mbox{P. Jacques\unskip,\iRUTG}
\mbox{J.A. Jaros\unskip,\iSLAC}
\mbox{Z.Y. Jiang\unskip,\iSLAC}
\mbox{A.S. Johnson\unskip,\iSLAC}
\mbox{J.R. Johnson\unskip,\iWISC}
\mbox{R.A. Johnson\unskip,\iCINC}
\mbox{T. Junk\unskip,\iSLAC}
\mbox{R. Kajikawa\unskip,\iNAGO}
\mbox{M. Kalelkar\unskip,\iRUTG}
\mbox{Y. Kamyshkov\unskip,\iTENN}
\mbox{H.J. Kang\unskip,\iRUTG}
\mbox{I. Karliner\unskip,\iILLI}
\mbox{H. Kawahara\unskip,\iSLAC}
\mbox{Y. D. Kim\unskip,\iSOGA}
\mbox{M.E. King\unskip,\iSLAC}
\mbox{R. King\unskip,\iSLAC}
\mbox{R.R. Kofler\unskip,\iMASS}
\mbox{N.M. Krishna\unskip,\iCOLO}
\mbox{R.S. Kroeger\unskip,\iMISSI}
\mbox{M. Langston\unskip,\iOREG}
\mbox{A. Lath\unskip,\iMIT}
\mbox{D.W.G. Leith\unskip,\iSLAC}
\mbox{V. Lia\unskip,\iMIT}
\mbox{C.Lin\unskip,\iMASS}
\mbox{M.X. Liu\unskip,\iYALE}
\mbox{X. Liu\unskip,\iUCSC}
\mbox{M. Loreti\unskip,\iPADO}
\mbox{A. Lu\unskip,\iUCSB}
\mbox{H.L. Lynch\unskip,\iSLAC}
\mbox{J. Ma\unskip,\iWASH}
\mbox{G. Mancinelli\unskip,\iRUTG}
\mbox{S. Manly\unskip,\iYALE}
\mbox{G. Mantovani\unskip,\iPERU}
\mbox{T.W. Markiewicz\unskip,\iSLAC}
\mbox{T. Maruyama\unskip,\iSLAC}
\mbox{H. Masuda\unskip,\iSLAC}
\mbox{E. Mazzucato\unskip,\iFERR}
\mbox{A.K. McKemey\unskip,\iBRUN}
\mbox{B.T. Meadows\unskip,\iCINC}
\mbox{G. Menegatti\unskip,\iFERR}
\mbox{R. Messner\unskip,\iSLAC}
\mbox{P.M. Mockett\unskip,\iWASH}
\mbox{K.C. Moffeit\unskip,\iSLAC}
\mbox{T.B. Moore\unskip,\iYALE}
\mbox{M.Morii\unskip,\iSLAC}
\mbox{D. Muller\unskip,\iSLAC}
\mbox{V.Murzin\unskip,\iMOSCOW}
\mbox{T. Nagamine\unskip,\iTOHO}
\mbox{S. Narita\unskip,\iTOHO}
\mbox{U. Nauenberg\unskip,\iCOLO}
\mbox{H. Neal\unskip,\iSLAC}
\mbox{M. Nussbaum\unskip,\iCINC}
\mbox{N.Oishi\unskip,\iNAGO}
\mbox{D. Onoprienko\unskip,\iTENN}
\mbox{L.S. Osborne\unskip,\iMIT}
\mbox{R.S. Panvini\unskip,\iVAND}
\mbox{C. H. Park\unskip,\iSOONG}
\mbox{T.J. Pavel\unskip,\iSLAC}
\mbox{I. Peruzzi\unskip,\iFRAS}
\mbox{M. Piccolo\unskip,\iFRAS}
\mbox{L. Piemontese\unskip,\iFERR}
\mbox{K.T. Pitts\unskip,\iOREG}
\mbox{R.J. Plano\unskip,\iRUTG}
\mbox{R. Prepost\unskip,\iWISC}
\mbox{C.Y. Prescott\unskip,\iSLAC}
\mbox{G.D. Punkar\unskip,\iSLAC}
\mbox{J. Quigley\unskip,\iMIT}
\mbox{B.N. Ratcliff\unskip,\iSLAC}
\mbox{T.W. Reeves\unskip,\iVAND}
\mbox{J. Reidy\unskip,\iMISSI}
\mbox{P.L. Reinertsen\unskip,\iUCSC}
\mbox{P.E. Rensing\unskip,\iSLAC}
\mbox{L.S. Rochester\unskip,\iSLAC}
\mbox{P.C. Rowson\unskip,\iCOLU}
\mbox{J.J. Russell\unskip,\iSLAC}
\mbox{O.H. Saxton\unskip,\iSLAC}
\mbox{T. Schalk\unskip,\iUCSC}
\mbox{R.H. Schindler\unskip,\iSLAC}
\mbox{B.A. Schumm\unskip,\iUCSC}
\mbox{J. Schwiening\unskip,\iSLAC}
\mbox{S. Sen\unskip,\iYALE}
\mbox{V.V. Serbo\unskip,\iSLAC}
\mbox{M.H. Shaevitz\unskip,\iCOLU}
\mbox{J.T. Shank\unskip,\iBU}
\mbox{G. Shapiro\unskip,\iLBL}
\mbox{D.J. Sherden\unskip,\iSLAC}
\mbox{K. D. Shmakov\unskip,\iTENN}
\mbox{C. Simopoulos\unskip,\iSLAC}
\mbox{N.B. Sinev\unskip,\iOREG}
\mbox{S.R. Smith\unskip,\iSLAC}
\mbox{M. B. Smy\unskip,\iCSU}
\mbox{J.A. Snyder\unskip,\iYALE}
\mbox{H. Staengle\unskip,\iCSU}
\mbox{A. Stahl\unskip,\iSLAC}
\mbox{P. Stamer\unskip,\iRUTG}
\mbox{H. Steiner\unskip,\iLBL}
\mbox{R. Steiner\unskip,\iADEL}
\mbox{M.G. Strauss\unskip,\iMASS}
\mbox{D. Su\unskip,\iSLAC}
\mbox{F. Suekane\unskip,\iTOHO}
\mbox{A. Sugiyama\unskip,\iNAGO}
\mbox{S. Suzuki\unskip,\iNAGO}
\mbox{M. Swartz\unskip,\iJHU}
\mbox{A. Szumilo\unskip,\iWASH}
\mbox{T. Takahashi\unskip,\iSLAC}
\mbox{F.E. Taylor\unskip,\iMIT}
\mbox{J. Thom\unskip,\iSLAC}
\mbox{E. Torrence\unskip,\iMIT}
\mbox{N. K. Toumbas\unskip,\iSLAC}
\mbox{T. Usher\unskip,\iSLAC}
\mbox{C. Vannini\unskip,\iPISA}
\mbox{J. Va'vra\unskip,\iSLAC}
\mbox{E. Vella\unskip,\iSLAC}
\mbox{J.P. Venuti\unskip,\iVAND}
\mbox{R. Verdier\unskip,\iMIT}
\mbox{P.G. Verdini\unskip,\iPISA}
\mbox{D. L. Wagner\unskip,\iCOLO}
\mbox{S.R. Wagner\unskip,\iSLAC}
\mbox{A.P. Waite\unskip,\iSLAC}
\mbox{S. Walston\unskip,\iOREG}
\mbox{J.Wang\unskip,\iSLAC}
\mbox{S.J. Watts\unskip,\iBRUN}
\mbox{A.W. Weidemann\unskip,\iTENN}
\mbox{E. R. Weiss\unskip,\iWASH}
\mbox{J.S. Whitaker\unskip,\iBU}
\mbox{S.L. White\unskip,\iTENN}
\mbox{F.J. Wickens\unskip,\iRAL}
\mbox{B. Williams\unskip,\iCOLO}
\mbox{D.C. Williams\unskip,\iMIT}
\mbox{S.H. Williams\unskip,\iSLAC}
\mbox{S. Willocq\unskip,\iMASS}
\mbox{R.J. Wilson\unskip,\iCSU}
\mbox{W.J. Wisniewski\unskip,\iSLAC}
\mbox{J. L. Wittlin\unskip,\iMASS}
\mbox{M. Woods\unskip,\iSLAC}
\mbox{G.B. Word\unskip,\iVAND}
\mbox{T.R. Wright\unskip,\iWISC}
\mbox{J. Wyss\unskip,\iPADO}
\mbox{R.K. Yamamoto\unskip,\iMIT}
\mbox{J.M. Yamartino\unskip,\iMIT}
\mbox{X. Yang\unskip,\iOREG}
\mbox{J. Yashima\unskip,\iTOHO}
\mbox{S.J. Yellin\unskip,\iUCSB}
\mbox{C.C. Young\unskip,\iSLAC}
\mbox{H. Yuta\unskip,\iAOMORI}
\mbox{G. Zapalac\unskip,\iWISC}
\mbox{R.W. Zdarko\unskip,\iSLAC}
\mbox{J. Zhou\unskip.\iOREG}

\it
  \vskip \baselineskip                   
  \vskip \baselineskip        
  \baselineskip=.75\baselineskip   
\iADEL
  Adelphi University, Garden City, New York 11530, \break
\iAOMORI
  Aomori University, Aomori , 030 Japan, \break
\iBOLO
  INFN Sezione di Bologna, I-40126, Bologna Italy, \break
\iBRI
  University of Bristol, Bristol, U.K., \break
\iBRUN
  Brunel University, Uxbridge, Middlesex, UB8 3PH United Kingdom, \break
\iBU
  Boston University, Boston, Massachusetts 02215, \break
\iCINC
  University of Cincinnati, Cincinnati, Ohio 45221, \break
\iCOLO
  University of Colorado, Boulder, Colorado 80309, \break
\iCOLU
  Columbia University, New York, New York 10533, \break
\iCSU
  Colorado State University, Ft. Collins, Colorado 80523, \break
\iFERR
  INFN Sezione di Ferrara and Universita di Ferrara, I-44100 Ferrara, Italy, \break
\iFRAS
  INFN Lab. Nazionali di Frascati, I-00044 Frascati, Italy, \break
\iILLI
  University of Illinois, Urbana, Illinois 61801, \break
\iJHU
  Johns Hopkins University, Baltimore, MD 21218-2686, \break
\iLBL
  Lawrence Berkeley Laboratory, University of California, Berkeley, California 94720, \break
\iLTU
  Louisiana Technical University - Ruston,LA 71272, \break
\iMASS
  University of Massachusetts, Amherst, Massachusetts 01003, \break
\iMISSI
  University of Mississippi, University, Mississippi 38677, \break
\iMIT
  Massachusetts Institute of Technology, Cambridge, Massachusetts 02139, \break
\iMOSCOW
  Institute of Nuclear Physics, Moscow State University, 119899, Moscow Russia, \break
\iNAGO
  Nagoya University, Chikusa-ku, Nagoya 464 Japan, \break
\iOREG
  University of Oregon, Eugene, Oregon 97403, \break
\iOXF
  Oxford University, Oxford, OX1 3RH, United Kingdom, \break
\iPADO
  INFN Sezione di Padova and Universita di Padova I-35100, Padova, Italy, \break
\iPERU
  INFN Sezione di Perugia and Universita di Perugia, I-06100 Perugia, Italy, \break
\iPISA
  INFN Sezione di Pisa and Universita di Pisa, I-56010 Pisa, Italy, \break
\iRAL
  Rutherford Appleton Laboratory, Chilton, Didcot, Oxon OX11 0QX United Kingdom, \break
\iRUTG
  Rutgers University, Piscataway, New Jersey 08855, \break
\iSLAC
  Stanford Linear Accelerator Center, Stanford University, Stanford, California 94309, \break
\iSOGA
  Sogang University, Seoul, Korea, \break
\iSOONG
  Soongsil University, Seoul, Korea 156-743, \break
\iTENN
  University of Tennessee, Knoxville, Tennessee 37996, \break
\iTOHO
  Tohoku University, Sendai 980, Japan, \break
\iUCSB
  University of California at Santa Barbara, Santa Barbara, California 93106, \break
\iUCSC
  University of California at Santa Cruz, Santa Cruz, California 95064, \break
\iUVIC
  University of Victoria, Victoria, B.C., Canada, V8W 3P6, \break
\iVAND
  Vanderbilt University, Nashville,Tennessee 37235, \break
\iWASH
  University of Washington, Seattle, Washington 98105, \break
\iWISC
  University of Wisconsin, Madison,Wisconsin 53706, \break
\iYALE
  Yale University, New Haven, Connecticut 06511. \break

\rm
%

\end{center}


\vfill
\eject

\vskip 1truecm

\vskip 1truecm
 
\end{document}